\documentclass{aa}
\usepackage{graphicx}
\usepackage{txfonts}
\usepackage{natbib}

\begin{document}

\title{A Sino-German $\lambda$6~cm polarization survey of the Galactic plane} 
\subtitle{I. Survey strategy and results for the first survey region}

\author{X. H. Sun\inst{1}
        \and J. L. Han\inst{1}
        \and W. Reich\inst{2} 
        \and P. Reich\inst{2}
        \and W. B. Shi\inst{1}
        \and R. Wielebinski\inst{2}
        \and E. F\"urst\inst{2}}

\offprints{X. H. Sun}

\institute{National Astronomical Observatories, Chinese Academy of
            Sciences, Jia-20 Datun Road, Chaoyang District, Beijing 100012, 
            China\\
            \email{xhsun,hjl,swb@bao.ac.cn}
            \and Max-Planck-Institut f\"{u}r Radioastronomie, 
                 Auf dem H\"ugel 69, 53121 Bonn, Germany\\
            \email{wreich,preich,rwielebinski,efuerst@mpifr-bonn.mpg.de}}

\date{Received / Accepted}

\abstract
{} 
{Polarization measurements of the Galactic plane at $\lambda$6~cm
probe the interstellar medium (ISM) to larger distances compared to
measurements at longer wavelengths, hence enable us to investigate
properties of the Galactic magnetic fields and electron density.}
{We are conducting a new $\lambda$6~cm continuum and polarization
survey of the Galactic plane covering $10\degr\leq l\leq230\degr$ and
$|b|\leq5\degr$. Missing large-scale structures in the $U$ and $Q$ maps 
are restored based on extrapolated polarization K-band maps from the 
WMAP satellite. The $\lambda$6~cm data are analyzed together with maps 
at other bands. 
}
{
We discuss some results for the first survey region, 
$7\degr\times10\degr$ in size, centered at $(l,b)=(125\fdg5, 0\degr)$. 
Two new passive Faraday screens, G125.6$-$1.8 and G124.9$+$0.1, were
detected. They cause significant rotation of background polarization angles
but little depolarization. G124.9$+$0.1 was identified as a new faint HII
region at a distance of 2.8~kpc. G125.6$-$1.8, with a size of about 46~pc, has 
neither correspondence in enhanced H$\alpha$ emission nor a counterpart in 
total intensity. A model combining foreground and background polarization 
modulated by the Faraday screen was developed. Using this model, we estimated 
the strength of the ordered magnetic field along the line of sight to be 
3.9~$\mu$G for G124.9$+$0.1, and exceeding 6.4~$\mu$G for G125.6$-$1.8. We 
obtained an estimate of 2.5 and 6.3~mK~kpc$^{-1}$ for the average polarized 
and total synchrotron emissivity towards G124.9$+$0.1. The synchrotron 
emission beyond the Perseus arm is quite weak.
A spectral curvature previously reported for SNR G126.2$+$1.6 is
ruled out by our new data, which prove a straight spectrum.
} 
{The new $\lambda$6~cm survey will play an important role in improving the 
understanding of the properties of the magneto-ionic ISM. The magnetic fields 
in HII regions can be measured. Faraday screens with very low electron 
densities but large rotation measures were detected indicating strong and 
regular magnetic fields in the ISM. Information about the local synchrotron 
emissivity can be obtained.
}

\keywords{Surveys -- Polarization -- Radio continuum: general -- Methods: 
observational -- ISM: magnetic fields}

\maketitle

\section{Introduction}

The first detection of diffuse polarized emission from the Milky Way
Galaxy \citep{wsb+62,wsp62} confirmed that its non-thermal radiation
originates in synchrotron emission. The two major sources of polarized
emission from our Galaxy are diffuse radio emission associated with
the Galactic disk produced by relativistic electrons spiraling in
interstellar magnetic fields, and discrete sources such as supernova
remnants (SNRs) with compressed interstellar magnetic fields, where
relativistic electrons are accelerated by shocks.

To understand the properties of the ISM in our Galaxy a number of
whole sky surveys \citep[e.g. as reviewed by][]{rei03} have been made.
Also large-scale radio surveys of the Galactic plane with higher
angular resolution were performed.  The large-scale surveys clearly
show the concentration of emission on the Galactic plane, which hosts
copious structures. Survey data at 408~MHz \citep{hssw82} were used to
construct a Galactic radio emission model \citep{bkb85}. The spectral
index distribution for the northern sky was determined by
\citet{rr88a}. A number of {\it polarization surveys} of our Galaxy
were carried out, as reviewed by \citet{rei06}. The milestone among
the early surveys is the multi-frequency mapping of the northern sky
by \citet{bs76} with the Dwingeloo 25\ m telescope at 408~MHz,
465~MHz, 610~MHz, 820~MHz and 1411~MHz. These surveys were absolutely
calibrated.

The Galactic plane polarization surveys experienced renaissance in the
1980s. A 2.7~GHz survey with 4\farcm3 resolution, conducted with
the Effelsberg 100\ m telescope, uncovered various patchy polarization
structures, most of which have no counterpart in total intensity
\citep{jfr87,rrf90,rfrr90,frrr90,drrf99}. Subsequent surveys like the
Parkes 2.4~GHz survey \citep{dshj95,dhjs97}, and the Effelsberg Medium
Latitude Survey (EMLS) at 1.4~GHz \citep{ufr+98,ufr+99,rfruww04}
continue to reveal diffuse polarization structures over various
scales. To achieve arcmin angular resolution, interferometers were
also used.  Large areas were observed at 350~MHz with the Westerbork
Synthesis Radio Telescope \citep[WSRT,][]{wbj+93,hkd03}. The Canadian
Galactic Plane Survey (CGPS) at 1.4~GHz was carried out with the
Dominion Radio Astrophysical Observatory~(DRAO) Synthesis Telescope
\citep{tgp+03,ulgr03}, and the Southern Galactic Plane Survey (SGPS)
at 1.4~GHz was conducted with the Australia Telescope Compact Array
\citep{gdm+01}.

Although a wealth of new data is available, we are still far away from
a clear picture of the ISM structure in our Galaxy, since the desired
information from the data is still limited. Significant depolarization
has been observed at low frequencies, which can be caused by intrinsic
ISM structures, e.g. random magnetic fields, within a telescope beam,
or Faraday effects from magnetized thermal gas either inside or in
front of emission regions.  The interferometer data have high angular
resolution to resolve details. The polarization maps show overwhelming
small-scale emission and depolarization features \citep[e.g. ``canals"
in ][]{hkd03} often interpreted as caused by fluctuations of the ISM
due to turbulent cells with scales comparable to observational beam
sizes. But the short-baseline data are missing so that large-scale
structures are not observed. Single dish observations pick up
large-scale structures, but small-scale structures cannot be
resolved due to coarser angular resolution.

Currently a Sino-German $\lambda$6~cm polarization survey of the
Galactic plane is carried out using the Urumqi 25~m radio telescope, which
has a resolution of $9\farcm5$, about the same as for the 1.4~GHz
EMLS.
These observations cover large regions and reveal large-scale
structures being missed in any synthesis telescope surveys at this
frequency. On the other hand, the Faraday effect is related with the
square of the observed wavelength \citep{tri91,sbs+98}, therefore,
Faraday depolarization is much smaller than that at lower frequencies,
and we can see much deeper into the ISM at 4.8~GHz. This is exactly 
the impetus of the $\lambda$6~cm polarization survey.

The new $\lambda$6~cm data are also quite valuable for studies of
large diameter SNRs, which cannot be observed easily by other
telescopes due to their large size in the sky and the high
sensitivity needed.  The continuum data can be used to investigate
whether there is a spectral curvature at high frequencies
\citep[e.g. that of S~147 by ][]{fr86}, which is important to
understand the late evolution of SNRs.  Due to little Faraday
modulation the polarization maps show the magnetic field structure of
SNRs very directly. They can also be used as probes for Faraday
tomography analysis to study the properties of the ISM both inside and
in the foreground of SNRs \citep{srh+06}.

The $\lambda$6~cm survey is also inspired by the current focus on
measurements of the cosmic microwave background polarization. The
synchrotron emission from our Galaxy is the major foreground
contamination, which has been modeled by various groups
\citep[e.g.][]{bcf+03}. The DRAO 1.4~GHz polarization survey of the
northern sky \citep{wlrw06}, our 4.8~GHz survey and the Wilkinson
Microwave Anisotropy Probe (WMAP) polarization data \citep{phk+06}
might be combined to yield a superb template.

Polarization data must be absolutely calibrated, since polarized
structures might be totally different in morphology after calibration
due to the non-linear dependence of the polarized intensity on the
Stokes parameter $U$ and $Q$ \citep[e.g.][]{rei06}. To obtain absolute
measurements of large-scale structures, the influence of the
environments (ground and atmosphere) and all kind of instrumental
effects \citep{wlrw06} have to be carefully removed.  In this paper,
the ground radiation and the instrument's effects are fit with a
first or second order polynomial, which is then subtracted from
the original data.
The lost large-scale structures are then recovered using the
K-band (22.8~GHz) data from WMAP \citep{phk+06} by spectral
extrapolation. This scheme is not really an absolute
calibration, but can be regarded to be sufficient for current analysis
of the $\lambda$6~cm data. Here, we present an extensive study of
the first region of the survey. In Section 2, the instrument and
survey strategy are briefly introduced. The data reduction is
described in Section 3. The survey map obtained is discussed in
Section 4.  A detailed study of individual objects is presented in
Section 5. Conclusions are summarized in Section 6.

\section{Instrumentation and survey observing strategy}

The Sino-German $\lambda$6~cm continuum and polarization survey of the
Galactic plane is being conducted using the Urumqi 25~m telescope
located at Nanshan station (87$\degr$~E, 43$\degr$~N) of the Urumqi
Observatory, National Astronomical Observatories of the Chinese
Academy of Sciences. The $\lambda$6~cm receiving system was
constructed at the Max-Planck-Institut f\"ur Radioastronomie (MPIfR)
in Germany and installed at the telescope in August 2004.  The survey
observations were started in September 2004.

The system has been briefly introduced by \citet{srh+06} and will be
detailed elsewhere (Reich et al., in prep.). The receiving system, in the
direction of the receiving process of signals, consists of: (1) a
corrugated feed installed in the secondary focus; (2) an orthogonal
transducer converting the signals into left-handed ($L$) and
right-handed ($R$) polarized components; (3) two cooled HEMT
pre-amplifiers working below 15~K in a dewar; (4) local oscillators;
(5) a polarimeter making the four correlations of $R$ and $L$
components ($LL^*$, $RR^*$, $RL^*$, $R^*L$); (6) voltage-frequency
converters converting the detected voltage signals to frequencies on
the antenna, which enables the long-distance transportation of the
signals to a digital backend in the control room; (7) a MPIfR ``Pocket
backend" in the control room that counts the frequency-coded signals from
the four channels. These raw data are transfered to a Linux PC and
stored on disk for further processing.

The backend can be conveniently remotely set from the control room for
the sampling time (i.e. the integration time for raw data)
and the duration of the injection of calibration signals. This system
is a copy from that used at the Effelsberg telescope
\citep{wlrm02}. One modulation cycle contains four phases.  The
sampling time (i.e. duration of one phase in modulation) is set to
32~msec as a standard, but can be changed to 16~msec if necessary. For
two subsequent phases within a cycle the calibration signal is
switched on, so that any gain changes of the system can be monitored.
The phase of the output signals is switched by $180\degr$
alternatively for every phase, that the quadratic terms of the
polarimeter are canceled. In one whole cycle, i.e. 128~ms, four
combinations of different settings of calibration and phase-switching
are realized.

The $\lambda$6~cm system has a system temperature about 22\ K, when
the telescope points to the zenith at clear sky. The half power beam
width (HPBW) is $9\farcm5$. The receiving system was designed for a
central frequency of 4800\ MHz and a bandwidth of 600\ MHz. However,
four groups of geostationary Indian satellites ({\it
InSat}) series located in southern direction emit strong signals
ranging up to 4810\ MHz. To suppress the interferences of the {\it
InSat}, a (tunable) filter was installed in November, 2005. The
receiver has two working modes now: a broad band mode for the northern
sky observations with a central frequency of 4800\ MHz, a bandwidth of
600\ MHz and a calibration signal of 1.7~K T$_{a}$; and a narrow band
mode with the central frequency of 4963\ MHz, a bandwidth of 295\ MHz
and a calibration signal of 1.45~K T$_{a}$.

The Sino-German $\lambda$6~cm polarization survey of the Galactic
plane is intended to map the Galactic plane within a range of Galactic
longitude (GL) of about $10\degr\leq l\leq230\degr$ and Galactic
latitude (GB) of $-5\degr\leq b\leq5\degr$. It is difficult to map
regions of smaller or larger longitudes, because such a region always
has an elevation at or below $10\degr$, where the ground radiation
contamination is very serious. Measurements of the Urumqi ground
radiation characteristics at $\lambda$6~cm were presented by
\citet{whss06}.

The survey regions are covered by raster scans in both GL and GB
directions, i.e. the sky is scanned at least twice.
The survey region is divided into fields covering $2\degr$ or
$2\fdg2\,({\rm GL})\times10\degr\,({\rm GB})$ for scanning in GB
direction, and a number of typically $8\degr\,({\rm
GL})\times2\fdg6\,({\rm GB})$ for scanning along GL direction, so that
each field can be observed in a reasonable time when the instrument or
other conditions (e.g. weather) are stable enough on average. We
usually use a scan velocity of $2\degr$/min. The separation between
two subscans is $3\arcmin$, which conforms to the Nyquist theorem and
therefore insures a full sampling.  An overlap of about $0\fdg1$
between the fields edges guarantees baselevel adjustment of the
neighbouring fields.  The length of the GL fields varies to avoid the
presence of strong emission structures at the boundaries.  All survey
observations are carried out during night time with clear sky to avoid
the influence of the solar emission via the far-sidelobes of the
telescope. 3C~286 and 3C~295 serve as primary polarized and
unpolarized calibrators respectively.  3C~138, 3C~48 and 3C~147 serve
as secondary calibrators. Calibrators are always observed before and
after the survey maps. The survey parameters are summarized in
Table~\ref{par_sur}.

\begin{table}[!htbp]
\caption{Survey parameters.}\label{par_sur}
\begin{tabular}{ll}\hline\hline
Parameters  &  Values \\\hline
System temperature & 22~K\\
Telescope beamwidth & 9\farcm5\\
SubScan separation & 3\arcmin\\
Scan velocity & 2\degr/min\\
Scan direction & GL and GB\\
Typical rms-noise for total intensity & 1.4~mK~T$_{\rm B}$\\
Typical rms-noise for $U$/$Q$         & 0.5~mK~T$_{\rm B}$\\
Typical rms-noise for $PI$            & 0.7~mK~T$_{\rm B}$\\
Central frequency & 4800~MHz/4963~MHz\\
Bandwidth & 600~MHz/295~MHz\\
Conversion factor T$_{B}$/S & 0.164~K/Jy\\\hline
\end{tabular}
\end{table}

\section{Survey data processing procedure}\label{data_proc}

In this section, we summarize the data processing steps. The raw data
from the ``Pocket backend'' as well as the telescope position read
from the telescope control PC are stored into a file in MBFITS format
\citep{mph05} on a Linux-PC for every frontend phase of 32\ msec.  For
each subscan, any gain drift of the receiving system is corrected and
a linear baseline is subtracted from the raw data. All subscans of a
field are combined into a map. We then remove bad subscans and
baseline distortions from the map, and finally
all observed maps of a field are averaged in a suitable way to obtain
the final map.  We show below that in our reduction pipeline the
ground radiation contamination has been largely removed.

\subsection{General procedure}

The raw data are stored for every subscan from the data flow of four
channels from the "Pocket backend", together with time information (in
MJD) and telescope position. We extract Stokes $I$, $U$ and $Q$ from
these data for each individual subscan. Details will be described
elsewhere (Reich et al. in prep.). We then arrange data from all
subscans to form maps of Stokes $I$, $U$ and $Q$.  The $U$ and $Q$
maps are then corrected for the parallactic angle, so that the
polarization angles are measured in the celestial coordinate
system. For the survey, we use the Galactic coordinate system, which
needs another transformation of $U$ and $Q$.  All maps are finally
transformed to the NOD2 format \citep{has74} so that we can adopt all
the mapping software developed at the MPIfR for further processing.
This has been already described by \citet{srh+06}.  First, the
baselines of some distorted subscans can be further corrected by a
second order polynomial fit, in case they are observed at low
elevations and a linear fit cannot remove the entire ground
radiation. Second, spiky interference or bad subscans are removed or
replaced by an interpolation of surrounding map pixels.
Often ``scanning effects" (stripes showing up along the
telescope driving direction) are still clearly seen in the $I$, $U$ 
and $Q$ maps, which is caused 
by system instabilities or static low-level interference, where the 
``unsharp masking'' method developed by \citet{sr79} is used to suppress 
these influences.

Positions of point sources in the maps are compared with those from
the NRAO VLA Sky
Survey\footnote{http://www.cv.nrao.edu/nvss/NVSSlist.shtml}
\citep[NVSS,][]{ccg+98}. Position differences are in general smaller
than $1\arcmin$. Occasionally, larger position offsets of the map
occur due to unidentified technical reasons, which then could be corrected by
shifting the map coordinates accordingly.

We calibrate the $I$, $U$ and $Q$ maps in respect to the primary
calibration source, 3C~286, which is assumed to have a flux density of
7.5~Jy at 4.8~GHz, a polarization angle of $33\degr$ and a
polarization percentage of 11.3\%. These data are taken from 
\citet{bgpw77} and \citet{ti80} and are consistent with Effelsberg 
calibration source observations.
For cleaning of instrumental polarization the measurements of the 
unpolarized calibrator 3C~295 are used. 

We then combine the processed maps of all sub-fields to obtain large
survey maps. The maps in the two orthogonal GL and GB directions were
first Fourier transformed and then added together in the Fourier
domain according to their weight as described by \citet{eg88}. This
method removes residual scanning effects. The spatial frequency map is
inversely transformed for the final intensity distribution map.

\subsection{Ground radiation}\label{gem}

Ground radiation is always picked up through the side-lobes of the
telescope. We have made intensive tests to measure the azimuth- and
elevation-dependent ground radiation characteristics of the Urumqi
station at $\lambda$6~cm band \citep{whss06}. For observations at high
elevations ($>30\degr$), the total intensity of the ground radiation
does not vary significantly with azimuth or elevation. Therefore, for
a typical survey map, the ground radiation adds a temperature gradient
to the map, which is sufficiently well subtracted by a polynomial fit
(usually of first order sometimes of second order) from the total
intensity channel for each subscan.

The elimination of ground radiation in the Stokes $U$ and $Q$ maps is
by no means trivial, but must be cleaned from the final survey
data. The ground radiation itself is not polarized in most directions.
However, the instrumental sidelobes are strongly polarized and
generate spurious polarization signals when they pick up emission
from the ground.  These signals can severely affect the
observation of very weak polarization signals on large scales. In
contrast to the basically linear variation for the total intensity
contamination with azimuth and elevation, the spurious polarization of
ground radiation is more complex and difficult to model.

For the first survey region, we tried a standard procedure for Stokes
$U$ and $Q$ maps. First, a linear fit is made using the data at the
two ends of a subscan, which is subtracted from the data. This step
has already been done prior to the raw NOD2 map
generation. Consequently, the polarization data in the $U$ and $Q$
maps are all relative to the ends of all subscans. By this procedure,
the linear components of the ground radiation in Stokes $U$ and $Q$
have been removed. The residual ground radiation, so far present,
varies with azimuth and elevation and thus manifests as obvious
stripes inclined to the scanning direction. The inclined stripes
can be similarly treated as scanning effects which can be virtually
suppressed by applying the ``unsharp masking" method \citep{sr79} and
the ``PLAIT" process \citep{eg88}, but require an appropriate rotation
of the map. After this procedure we are confident that the spurious
polarization has been completely removed from the Stokes $U$ and $Q$
maps.

\subsection{Polarization Cleaning}
Based on observations of the unpolarized calibrators 3C\ 295 and 3C\
147, the instrumental $U$ and $Q$ have been measured showing a
``butterfly''-shaped symmetric configuration. Therefore the
instrumental $PI$ manifests as ring-like structures with a peak
percentage polarization of up to about 2\%.  The instrumental
polarization can be safely ignored except for strong sources or
Galactic structures, where a clean procedure should be applied. By
averaging a number of observations of 3C\ 295 and 3C\ 147 we obtained
Stokes $I$, $U$ and $Q$ maps of the instrumental response, which can
be regarded as instrumental pattern from the antenna and leakage from
$I$ into the polarization channels. The "REBEAM" procedure from the
NOD2 package is then applied \citep[e.g. ][ for more details see Reich
et al. in prep.]{sof87}.  In brief, for an object with observed Stokes
$I_{\rm obs}$, $U_{\rm obs}$ and $Q_{\rm obs}$, the instrumental
contribution to polarization ($U_{\rm inst}$ and $Q_{\rm inst}$) can
be obtained as, $\tilde{U}_{\rm inst}=\tilde{I}_{\rm
obs}\tilde{U}/\tilde{I}$ and $\tilde{Q}_{\rm inst}=\tilde{I}_{\rm
obs}\tilde{Q}/\tilde{I}$, where the tilde means the Fourier
transformation. The inverse transformation yields the instrumental
contribution of $U_{\rm inst}$ and $Q_{\rm inst}$, which is to be
subtracted from the $U_{\rm obs}$ and $Q_{\rm obs}$ maps to get clean
maps.

\begin{figure}[!htbp]
\includegraphics[width=0.45\textwidth]{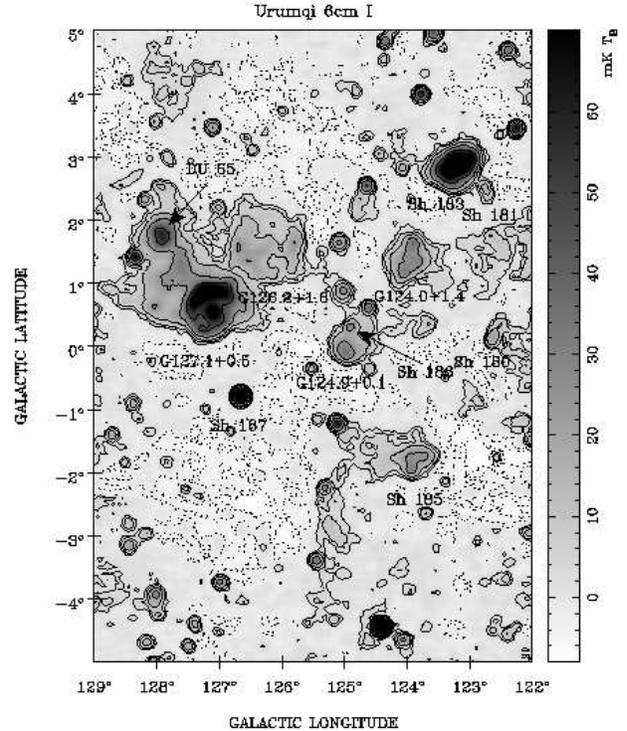}
\caption{The gray-scale image shows the total intensity for the survey
field centered at $(l, b)=(125\fdg5, 0\degr)$ with a size of
$7\degr\times10\degr$. The contours encode the total intensity with
levels equal to $\pm2^n\times3\sigma_I$ with $n=0,1,\dots$ and
$\sigma_I$ = 0.85~mK T$_{B}$. Solid lines show positive
intensities and dotted lines negative ones. 
}
\label{6cmi}
\end{figure}

\section{Observation and processing of the first survey region}

In this paper we analyse objects from the first survey region centered
at $(l, b)=(125\fdg5, 0\degr)$ with a size of $7\degr\times10\degr$.
Observations for two coverages in the GL direction and two coverages
in the GB direction were conducted between October 2004 and April
2006.

Following the data processing procedure described in Section
\ref{data_proc}, we obtained the total intensity map as shown in
Fig.~\ref{6cmi}, the Stokes $U$ and $Q$ maps in Fig.~\ref{6cmuq} and
the polarized intensity and polarization angle maps in
Fig.~\ref{6cmpipa}. All these maps are on a relative level with the
edges set to zero. Therefore the large-scale structures comparable to
the map size are missed, which introduces a non-linear bias spurious
for the polarization results \citep{rei06}.

\subsection{Zero-level restoration}
\begin{figure*}[!htbp]
\begin{minipage}[t]{0.5\textwidth}
\centering
{\bf Original $U$}
\end{minipage}
\begin{minipage}[t]{0.5\textwidth}
\centering
{\bf Restored $U$}
\end{minipage}
\begin{minipage}{\textwidth}
\includegraphics[width=0.5\textwidth]{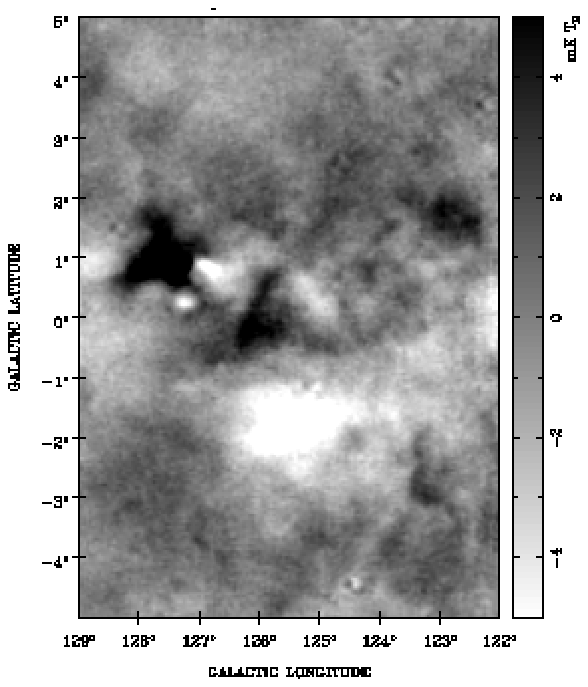}
\hspace*{\fill}
\includegraphics[width=0.5\textwidth]{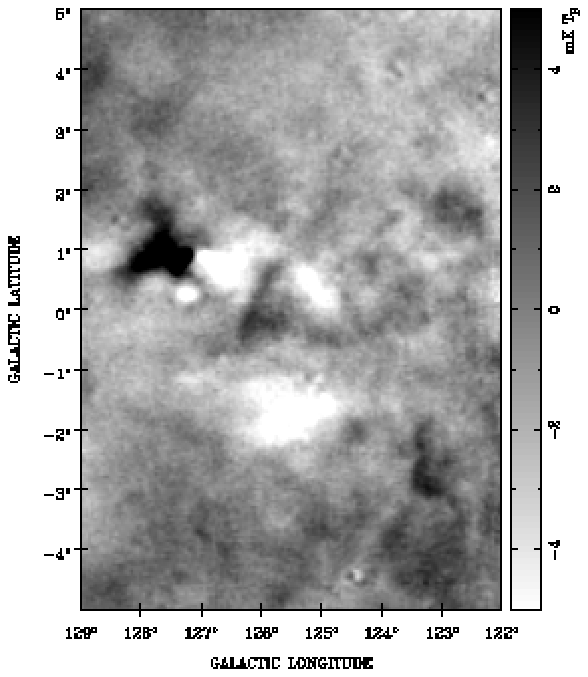}
\end{minipage}\vspace{0.2cm}
\begin{minipage}[t]{0.5\textwidth}
\centering
{\bf Original $Q$}
\end{minipage}
\begin{minipage}[t]{0.5\textwidth}
\centering
{\bf Restored $Q$}
\end{minipage}
\begin{minipage}{\textwidth}
\includegraphics[width=0.5\textwidth]{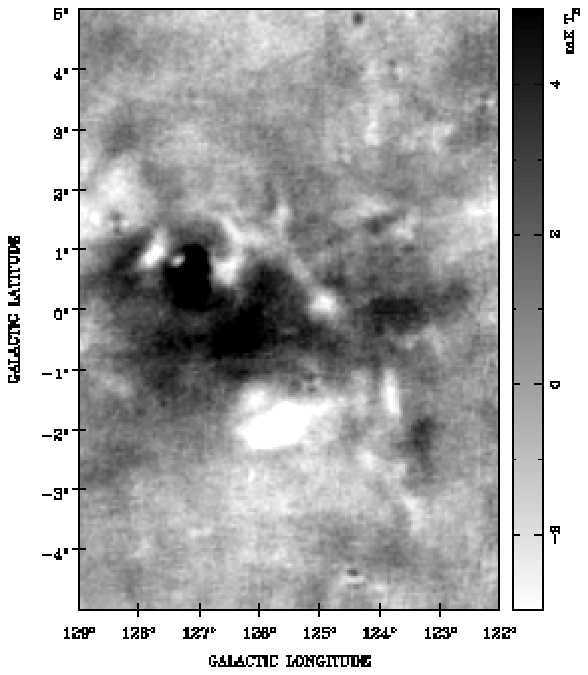}
\hspace*{\fill}
\includegraphics[width=0.5\textwidth]{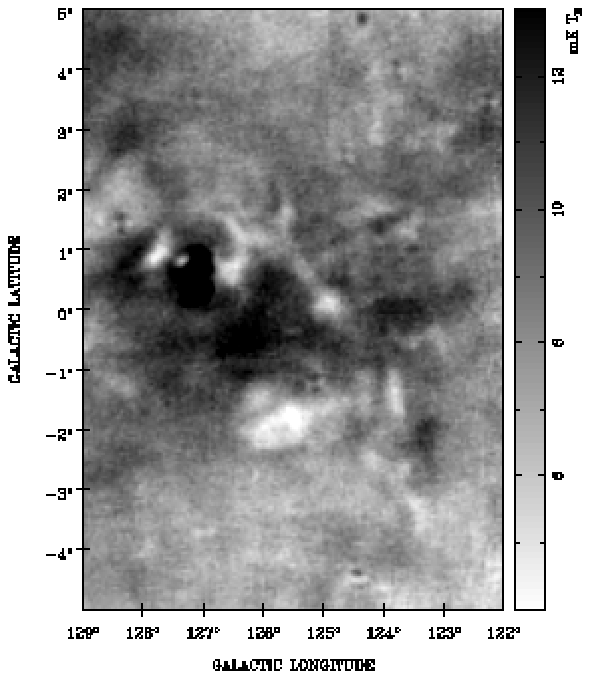}
\end{minipage}
\caption{Gray-scale images of the original and restored $U$ and $Q$ maps.}
\label{6cmuq}
\end{figure*}

\begin{figure*}[!htbp]
\begin{minipage}[t]{0.5\textwidth}
\centering
{\bf Original PI + B-vectors}
\end{minipage}
\begin{minipage}[t]{0.5\textwidth}
\centering
{\bf Restored PI + B-vectors}
\end{minipage}
\begin{minipage}{\textwidth}
\includegraphics[width=0.5\textwidth]{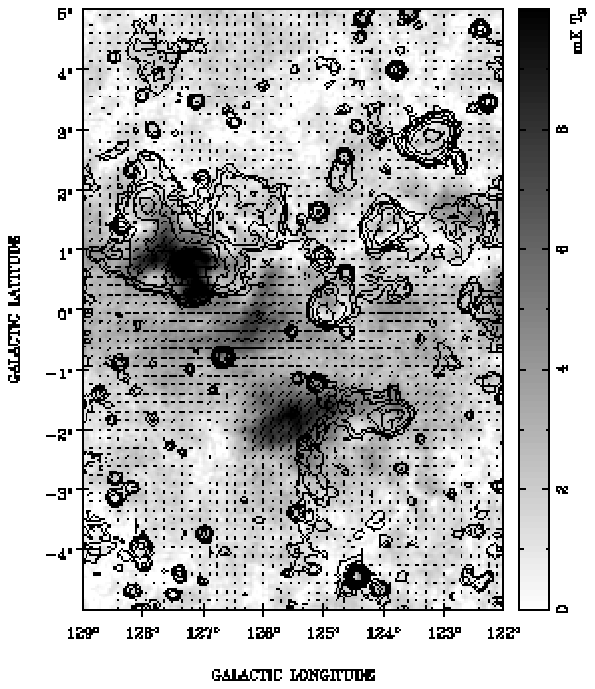}
\includegraphics[width=0.5\textwidth]{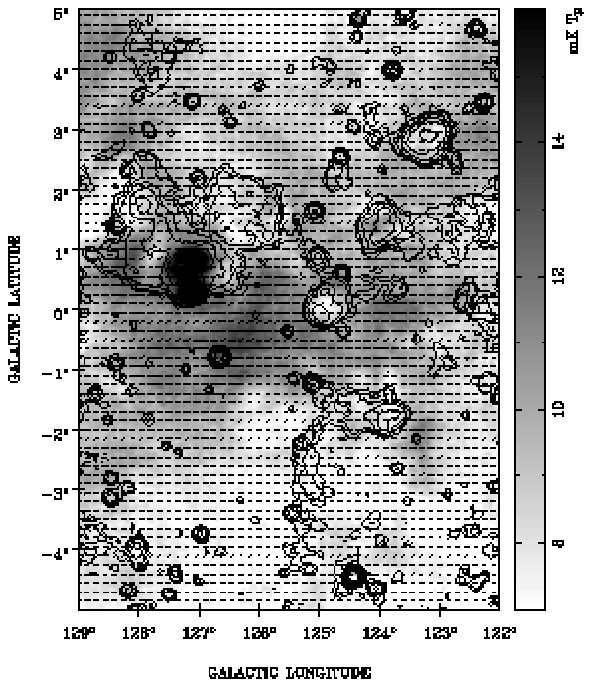}
\end{minipage}
\caption{Gray-scale images of the polarized intensity ($PI$) with
overlaid bars for every third pixels in B-field direction (polarization angle $PA+
90\degr$). The length of the bars is proportional to $PI$ with a lower
limit of about $\sigma_{PI}=0.4$~mK~T$_{\rm B}$. A polarized
intensity of 1~mK~T$_{\rm B}$ corresponds to a bar-length of
0.2 degree. Contours show total intensities with 
the same (positive) levels as in Fig.~\ref{6cmi}.}
\label{6cmpipa}
\end{figure*}

Most of the large-scale polarized emission missed in our field is
probably originating from the so-called ``Fan region'', which is an
outstanding strong polarization feature and can be easily recognized
from all existing polarization survey maps
\citep[e.g.][]{bs76,wlrw06}. \citet{ws74} and \citet{spo84} claimed
that the ``Fan region'' is a local feature at a distance of about
500~pc and Faraday rotation is negligible. We assume that the ``Fan
region'' is fully included in the 22.8~GHz (K-band) polarization map
from WMAP \citep{phk+06} which thus can be used for an estimate of
missing large-scale polarization components in our $\lambda$6~cm map.
We tried to compensate the missing large-scale structures in the $U$
and $Q$ maps in a similar way as described in \citet{ufr+98}.  First,
we convolved both our Urumqi $\lambda$6~cm map and the corresponding
K-band $U$ and $Q$ maps to a HPBW of $2\degr$. Second, we scaled the
smoothed K-band maps by the factor of $(\frac{4.8}{22.8})^\beta$, with
a spectral index $\beta$ of $-2.8$, assuming it is the same as that
for total intensities obtained by \citet{rr88a,rr88b}. We note that
the polarized intensity for the large-scale structures is about 300~mK
in the DRAO 1.4~GHz map \citep{wlrw06} and about 112~$\mu$K in the
K-band map, which yields the same spectral index of about $-$2.8.
Third, we subtracted the convolved Urumqi $\lambda$6~cm maps from the
scaled and smoothed K-band maps to obtain the difference
maps. Finally, we added the difference to the original Urumqi $U$ and
$Q$ maps.  Based on such zero-level restoration for $U$ and $Q$
maps, the polarized intensity and polarization angle maps were
recalculated (Figs.~\ref{6cmuq} and \ref{6cmpipa}).  The polarized
intensity of the missing large-scale structures is about 8.5~mK.  As
expected from the polarization angle of the ``Fan region'' , which is known
to be around zero, the zero-level correction of the $Q$ map is large, about 
8.3~mK, while that of $U$ map about $-0.6$~mK.

\begin{figure}[!htbp]
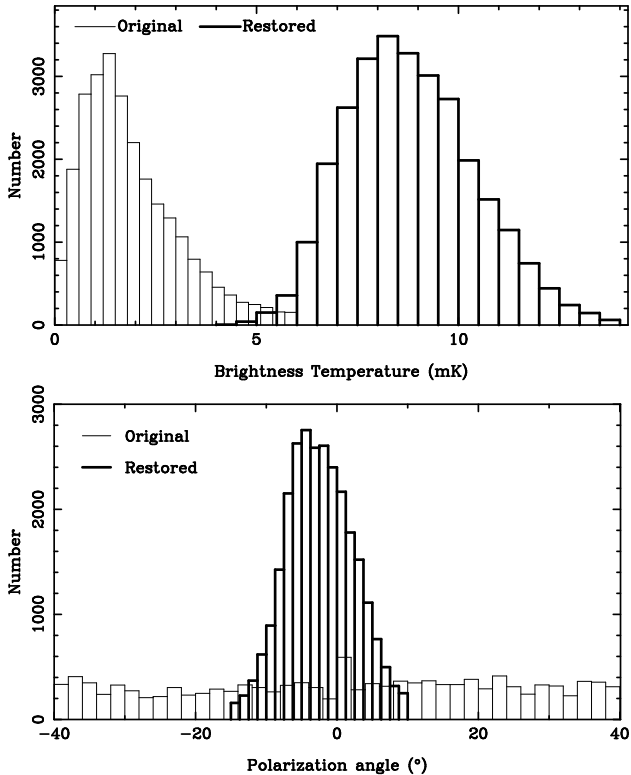

\includegraphics[bb=135 43 534 700,clip,angle=-90,width=0.45\textwidth]{pi_hist.ps}
\includegraphics[bb=116 43 535 708,clip,angle=-90,width=0.45\textwidth]{pa_hist.ps}
\caption{The distribution of polarized intensity and polarization
angles before and after restoration of large-scale structures 
from convolved K-band $U$ and $Q$ maps as observed by WMAP
(see section 4).}\label{pipad}
\end{figure}

The $U$ and $Q$ maps with the restored large-scale structures have to be
regarded as an approximation, which, however, is not too
far from the real situation. Compared to the observed $U$ and $Q$
maps, the structures change dramatically after the zero-level
restoration. For polarized intensity, a large-scale offset of about
8~mK has been added (Fig.~\ref{pipad}).  The most striking difference
can be found in the region at $(l, b)=(125\fdg5, -2\fdg2)$, where the
enhanced polarized intensity in the original maps is significantly
reduced. The change of the polarization angle distribution is obvious
from Figs.~\ref{6cmpipa} and \ref{pipad}. The polarization angles are
much less scattered when the large-scale components are included (see
Fig.~\ref{pipad}). The width of the polarization angle distribution
and the mean level of polarized intensities depends on the amount of
large scale emission added to the original data. Errors in the assumed
spectral index used to extrapolate the WMAP polarization data towards
$\lambda$6~cm have an effect on that, but the principal structure
and variations are preserved compared to the original maps.

\subsection{Accuracy and errors}
The typical rms noise for total intensity, $U$ and $Q$, and polarized
intensity were obtained from survey maps where the contribution from
Galactic structures are nearly absent. The results are listed in
Table~\ref{par_sur}.  

Because the system temperature $T_{\rm sys}$ is
22~K, the bandwidth $\Delta\nu$ is 600~MHz and the integration time
$\tau$ is 2 seconds, the equation $\sigma_I=T_{\rm
sys}/\sqrt{\Delta\nu\tau}$ gives the rms noise ($\sigma_I$) of 0.6~mK 
for total intensity. This corresponds to a brightness temperature of 
0.9~mK, which is calculated using the measured beam efficiency of 67\%. 
Since $U$ and $Q$ are
measured by correlating the two total intensity channels, their rms
noise is lower by a factor of $\sqrt{2}$ than that of the total
intensity, which means about 0.7~mK. For the first region discussed 
here, we have made two additional coverages with an integration time of 
1.5 seconds for each and one additional coverage with standard integration 
time of 1 second. The theoretical rms noise is now 0.6~mK~T$_{\rm B}$ 
for total intensity and 0.4~mK~T$_{\rm B}$ for $U$ and $Q$.
The measured rms noise for $I$ is 0.85~mK~T$_{\rm B}$ and 
for $U/Q$ is 0.3~mK~T$_{\rm B}$. For polarization, the prediction 
is consistent with the measurements. For total intensity, the measured 
rms noise is slightly higher, which is likely due to limited system 
stability and low-level interference. 
 
Our long-term observations of the primary survey calibrator 3C~286  
(from August, 2004 to April, 2006) show that the systematic error for 
total intensity calibration is less than 4\% and less than 5\% for 
polarized intensity. The observations also yield the  
polarization angle of $32\degr\pm1\degr$ for 3C~286, which is rather stable 
and very close to the assumed standard value. So we 
do not make further correction for the polarization angles and quote the 
uncertainty of the angles as 1$\degr$ for high signal-to noise ratios. 
We also obtained the 
flux density, polarization angle and polarization percentage for the 
secondary polarized calibrators 3C~48 and 3C~138.  
For 3C\ 48, these quantities are 5.5$\pm$0.1~Jy, $108\degr\pm1\degr$, 
and (4.2$\pm$0.4)\%. For 3C\ 138, the three quantities are
3.9$\pm$0.1~Jy, $169\degr\pm1\degr$ and (10.8$\pm$0.5)\%. The results are 
consistent with the values quoted in \citet{bgpw77} and \citet{ti80}. 
3C\ 48 and 3C\ 138 are known for slight variations with time.

To check whether the ground radiation has been fairly removed, we 
compared the final map with each individual map and found that 
all the common structures are preserved. This means that the ground 
radiation, if any still remains in the results at all, is below the 
level of rms noise. This can result in slightly larger rms noise than 
theoretical prediction as shown above.  

For zero-level restoration, we added just the large-scale $U$ and 
$Q$ components to the original maps, which do not introduce extra noise. 
However, the spectral index for polarized intensity is uncertain. 
If the spectral index varies by 0.1, the level of the restored polarized 
intensity will change by about 17\%. This means the peak of the polarized 
intensity distribution after zero-level restoration (Fig.~\ref{pipad}) will 
shift towards the lower or higher values. 
The polarization angle distribution will also change, depending how the
large scale components distribute in $U$ and $Q$. For our region the 
missed offsets in $Q$ are much larger than in $U$. 
Thus spectral index errors will 
change the width of the angle distribution and slightly shift the
mean absolute value. These shifts will, however, only slightly 
affect the polarization structures, and have little influence on our 
results on Faraday screens because our results are based on the difference 
or ratio between the polarized emission towards the screen and the emission  
in the surroundings. However, a spectral index error affects the emissivity.  

\section{Extended objects in the first survey region}

In the first survey region we recognize a number of compact sources and 
extended sources or structures. A detailed 
analysis of compact sources will be given elsewhere (Shi et al., in 
preparation). Here we analyze the extended sources.

The prominent extended objects seen in the total intensity map
(Fig.~\ref{6cmi}) are listed in
Table~\ref{objects}, together with their distances - where known -
and references.  The objects are SNRs, HII regions or reflection
nebulae. Two SNRs are identified according to the catalog of
\citet{gre06}, seven known HII regions (Sharpless HII regions and
DU~65) are identified from the catalog by \citet{sha59} and
\citet{dc76}. Four so far uncatalogued objects in our map are listed
at the end of the table. G128.4$+$4.3 and G122.7$+$1.5 are too weak at
other bands, so we did not explore their properties further.

Polarization maps are directly related to the magnetic field
structure. After adding the large-scale structures, the polarization
angles are concentrated around $0\degr$ (Figs.~\ref{6cmpipa} and
\ref{pipad}), indicating a very uniform large-scale magnetic field
running parallel to the Galactic plane. Some prominent features can
also be recognized from the polarized intensity and B-vector
maps~(Fig.~\ref{6cmpipa}), including polarization minima and regions
with polarization angles considerably deviating from the general
tendency. Many of these polarized structures have no counterpart in
total intensity.

The salient extended features in Figs.~\ref{6cmi}-\ref{6cmpipa} are: 
\begin{itemize}
\item SNRs G126.2$+$1.6 and G127.1$+$0.5, which have polarization
      detected in the relative scale maps. However the polarization
      towards the eastern shell of SNR G126.2$+$1.6 is considerably
      reduced after the zero-level restoration of $U$ and $Q$ maps.
\item A polarization feature at $(l,b)=(126\fdg2,-0\fdg2)$ seen
      in the original PI map in Fig.~\ref{6cmpipa}, which has no
      correspondence in total intensity and virtually disappears in 
      the restored map.
\item The strong polarized feature at $(l,b)=(125\fdg6,-1\fdg8)$ seen in the
      original PI map without a counterpart in the total intensity
      map. After the restoration, the polarized intensity is slightly below
      that of its surroundings, however, the polarization angles
      remarkably deviate from the surroundings.
\item The extended source G124.9$+$0.1, which shows polarization in
      the relative map in Fig.~\ref{6cmpipa}, but turns into a
      polarization minimum after the zero-level restoration.
\item The reflection nebula Sh~185, which is probably physically
      connected to the shell extending towards southeastern
      direction. In the zero-level restored map in Fig.~\ref{6cmpipa}, the
      polarization properties towards Sh~185 are unchanged, which
      means that the nebula does not emit polarized emission nor
      depolarizes or causes Faraday rotation of the background emission.
\item Polarization minima towards extended sources, such as
      Sh~183, are seen in both the original and the zero-level restored
      polarization maps, although the appearance might be patchy.
\end{itemize}

\begin{table*}[!htbp]
\caption{Prominent extended objects in total intensity map at $\lambda$6~cm.}
\begin{center}
\label{objects}
\begin{tabular}{lcrccll}
\hline\hline
Name & $l$ & $b$~ & $S_{4800}$ & distance   & Note & Ref. for distance\\ 
     & ($\degr$) & ($\degr$) & (Jy) & (kpc) & & \\
\hline
G127.1$+$0.5 & 127.1~ & 0.5~~  & 6.3$\pm$0.7 &  1.2, 4.4  & SNR         & 1, 2\\
G126.2$+$1.6 & 126.2~ & 1.6~~  & 2.6$\pm$0.6 &  2.4, 5.6  & SNR         & 2, 3\\
DU~65        & 127.9~ & 1.7~~  &             &   3        & HII region  & 4   \\
Sh~180       & 122.63 & 0.05   &             &  6.2       & HII region  & 5   \\
Sh~181       & 122.72 & 2.37   &             &  2.8       & HII region  & 6   \\
Sh~183       & 123.28 & 3.03   & 5.4$\pm$0.5 & 0.7, 6.2   &HII region   & 6,7,13 \\ 
Sh~185       & 123.85 & $-$1.97& 1.4$\pm$0.5 &   0.2      &Reflection nebula &8, 9, 10 \\  
Sh~186       & 124.90 & 0.32   &             & 2, 3.5     & HII region & 11, 5\\
Sh~187       & 126.66 & $-$0.79& 1.2$\pm$0.1 &  1         & HII region & 12 \\
G124.9$+$0.1 & 124.90 & 0.10   & 1.4$\pm$0.2 &  2.8       & HII region & 13 \\
G124.0$+$1.4 & 123.95 & 1.40   & 1.6$\pm$0.3 &            & HII region & 13 \\
G124.8$+$4.3 & 124.8~ & 4.3~~  &             &            &            & 13 \\
G122.7$+$1.5 & 122.7~ & 1.5~~  &             &            &            & 13 \\
\hline
\end{tabular}\\
\end{center}
{\footnotesize Reference for the distance: 
1. \citet{lt06}; 2. \citet{jrd89};
3. \citet{tl06}; 4. \citet{cac+03};
5. \citet{bfs82}; 6. \citet{fb84};
7. \citet{larv92}; 8. \citet{plk+97};
9. \citet{knm05}; 10. \citet{bmd+97}; 
11. \citet{cp03}; 12. \citet{jdr92}; 13. This paper.
}
\end{table*}

\subsection{Flux density and spectral index}\label{ana_tp}

To achieve an integrated net flux density of an extended source the
contributions of the diffuse background and extragalactic sources must
be removed. In this paper, the ``background filtering" technique
developed by \citet{sr79} is applied to subtract the large-scale
diffuse emission. The filtering beam was taken to be
$33\arcmin\times33\arcmin$, which gives roughly the scale length of
the separation between large-scale and small-scale emission.  Then a
"twisted" hyper-plane fitted by using the pixel values surrounding the
source is subtracted. In order to figure out the contribution from
compact sources, we extracted all point sources towards the target
object from the NVSS source catalog \citep{ccg+98}. The total flux
density of these sources is then extrapolated to the frequency of
4.8~GHz from 1.4~GHz with a spectral index of $\alpha=-0.8$
($S_\nu\propto\nu^\alpha$ with $S_\nu$ being the flux density at a
frequency $\nu$) to yield the source contribution.  The uncertainty of
the background level and the source contribution together introduce a
typical error of less than 10\% of the flux density of an object.

\begin{table*}[!htbp]            
\begin{center}
\caption{Spectral index $\beta$ of extended objects derived
from TT-plots. Note that the CGPS 1420 MHz data include the large-scale 
component from the Effelsberg 1408 MHz survey.  The average of
$\alpha$ is also listed.}
\label{tt_par}
\begin{tabular}{lcr@{$\pm$}lr@{$\pm$}lr@{$\pm$}lr@{$\pm$}lr@{$\pm$}lr}
\hline\hline
Region & radius &
         \multicolumn{2}{c}{$\beta_{4800/1420}$}&
         \multicolumn{2}{c}{$\beta_{4800/1408}$}& 
         \multicolumn{2}{c}{$\beta_{4800/865}$} & 
         \multicolumn{2}{c}{$\beta_{4800/408}$} &
         \multicolumn{2}{c}{average $\alpha$}   & $R_{60\mu m / 6cm}$\\
\hline
SNR G127.1$+$0.5 & 30\arcmin & $-$2.41&0.01 & $-$2.40&0.01 & $-$2.47&0.03         & $-$2.42&0.01 & $-$0.41&0.02 &  280 \\
SNR G126.2$+$1.6 & 40\arcmin & $-$2.58&0.05 & $-$2.47&0.04 & $-$2.55&0.05         & $-$2.51&0.09 & $-$0.52&0.03 &  770 \\
DU 65            & 24\arcmin & $-$1.91&0.02 & $-$1.91&0.02 & $-$2.07&0.09         & $-$2.01&0.28 &    0.09&0.07 &  440 \\
Sh~183           & 27\arcmin & $-$2.02&0.01 & $-$2.06&0.01 & \multicolumn{2}{c}{} & $-$1.90&0.02 & $-$0.02&0.01 &  420 \\
Sh~185           & 24\arcmin & $-$2.01&0.06 & $-$2.03&0.07 & \multicolumn{2}{c}{} & $-$2.12&0.14 & $-$0.03&0.06 & 1500 \\
G124.9$+$0.1     & 27\arcmin & $-$1.97&0.10 & $-$2.01&0.10 & \multicolumn{2}{c}{} & $-$2.15&0.36 &    0.00&0.13 & 1100 \\
G124.0$+$1.4     & 30\arcmin & $-$2.20&0.08 & $-$2.20&0.06 & \multicolumn{2}{c}{} & $-$2.34&0.11 & $-$0.22&0.05 &  600 \\
Sh~187           & 5\arcmin &                    \multicolumn{10}{c}{}                                         & 5800 \\
\hline
\end{tabular}
\end{center}
\end{table*} 

The spectral index can be obtained by fitting a power-law to the
integrated flux densities observed at various frequencies. However,
the spectral index could be influenced by the uncertainty of the
background level. For example, a background level uncertainty of 10\%
at 4.8~GHz can introduce an error of about $-$0.1 for the spectral
index between 4.8~GHz and 1.4~GHz. Therefore, we also obtained the
spectral index for the brightness temperature $\beta$ via
temperature-temperature plots~(TT-plots), which are unaffected by the
uncertainty of the background level. The flux density and brightness
temperature are related via $S_\nu\propto\nu^2T_\nu$, so the spectral
index for brightness temperatures can be translated to the spectral
index for flux densities as $\alpha=\beta+2$.  Fortunately many survey
data are public and thus facilitates the study of TT-plots. We
retrieved the CGPS 1420~MHz and 408~MHz survey
 data\footnote{http://www2.cadc-ccda.hia-iha.nrc-cnrc.gc.ca/cgps}, which 
includes Effelsberg data for a correct representation of 
the large-scale emission, 
and also the Effelsberg 1408~MHz survey data 
separately\footnote{http://www.mpifr-bonn.mpg.de/survey.html}. The published
865~MHz data around SNRs G127.1$+$0.5 and G126.2$+$1.6 by
\citet{rzf03} were also used. We convolved all data to a common
HPBW of $10\arcmin$ except for the 865~MHz data
and then obtained the spectral indices $\beta$ and
$\alpha$ listed in Table~\ref{tt_par}. For the spectral index between our 
4800~MHz data and the 865~MHz data, we smoothed the 4800~MHz data to a HPBW of 
$14.5\arcmin$.
Note that strong point sources
have been removed before TT-plots were made. 

The spectral index is an important diagnostic tool to determine the
nature of an extended source. An object with a spectral index around
$-$0.5 identifies that source as a SNR~(Table.~\ref{tt_par}). However,
a flat spectrum could originate from either a thermal source such as a
HII~region or a nonthermal source such as a plerion or a Crab-like
SNR. The identification of a HII~region can be strengthened if the
exciting star can be found. Below we have intensively used the
SIMBAD\footnote{http://simbad.u-strasbg.fr/Simbad} data base for the
search of exciting stars.

\begin{figure}
\includegraphics[width=0.45\textwidth]{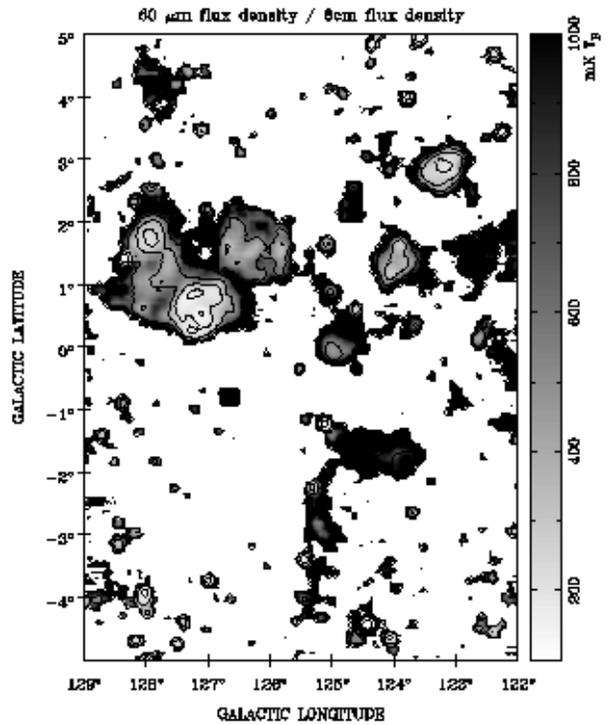}
\caption{The ratio of the IRAS 60~$\mu$m flux density to $\lambda$6~cm
flux density. The lower limit for the $\lambda$6~cm intensity is
3$\sigma_I$. Both maps were scaled to Jy/beam at
the common beamwidth of $10\arcmin$ before the ratio was calculated.
The superposed contours are at the same levels (above $3\sigma_I$) as
in Fig.~\ref{6cmi}.}
\label{060_6cm}
\end{figure}

Another criterion is the ratio between the infrared flux density and
the radio continuum flux density. \citet{frs87} have reported that the
ratio of the IRAS 60~$\mu$m intensity and the $\lambda$11~cm intensity
$R_{\rm 60\,\mu m/11\,cm}$ is around $500\lesssim R_{\rm 60\,\mu
m/11\,cm}\lesssim1500$ for HII~regions and $R_{\rm 60\,\mu
m/11\,cm}\lesssim250$ for SNRs for objects in the first Galactic
quadrant.  This reflects that HII~regions are strong infrared
emitters.  Therefore we calculated the ratio $R_{\rm 60\,\mu m/6\,cm}$
to determine the nature of a source. We retrieved the high resolution
IRAS 60~$\mu$m data \citep{ctpb97} from the CGPS data archive. We
scaled both the Urumqi total intensity data and the IRAS data to units
of Jy/beam.  The data were convolved to $10\arcmin$ and the ratio was
calculated as shown in Fig.~\ref{060_6cm}. The average of the ratios
for some extended objects are listed in the last column of
Table~\ref{tt_par}. The ratio for the SNR G127.1$+$0.5 is obviously
smaller than those for the other nebulae, but that for SNR
G126.2$+$1.6 is rather large, probably due to its very weak surface
brightness and enhanced large-scale infrared emission in its
direction. As can be seen from Fig.~\ref{060_6cm}, except for the
known SNRs, all extended sources all show very large ratio $R_{\rm
60\,\mu m/6\,cm}$, which indicates their thermal natures. No new SNR
or plerion could be detected in the first survey region.

\subsection{Interpretation of polarization structure}\label{ana_pi}

The observed polarization structures could be produced by two
mechanisms. Synchrotron emission is intrinsically polarized, and the
polarized intensity depends on the amount of the regular magnetic
field component in the emitting volume. In general SNRs are polarized 
objects, while HII~regions are not. However, as we show below, this 
is not necessarily directly observed in maps of polarized emission.  
A polarized structure can also be caused by Faraday effects within the
diffuse foreground ISM or a HII~region along the line of sight.
The polarization in the
direction of a source could be the sum of modulated
background and foreground components, which can be recognized only
from the zero-level restored maps.

\subsubsection{The polarization of radio emission}

As can be seen from Fig.~\ref{6cmpipa}, large-scale coherent polarized
structures prevail in both the original relative and zero-level restored
maps.  These structures should originate either in the Perseus arm of
about 2~kpc distance \citep{xrzm06} or in the ISM up to the Perseus
arm. As we show below, the polarized emission behind the Perseus arm 
is very weak. 
\citet{bkb85} modeled the synchrotron emissivity of about
11~K~kpc$^{-1}$ at 408~MHz. This corresponds to an emissivity of
11~mK~kpc$^{-1}$ at 4.8~GHz based on a spectral index of $-$2.8.
Depolarization is quite small at 4.8~GHz in case there are no Faraday
screens along the line of sight. Thus the polarization percentage can
be written as $p=p_i B_{\rm reg}^2/(B_{\rm reg}^2+B_{\rm ran}^2)$,
where $p_i\approx75\%$ is the intrinsic polarization percentage, and
$B_{\rm reg}$ and $B_{\rm ran}$ are the regular and random magnetic
field components, respectively. Providing a distance of 2--3~kpc and a
polarization percentage of about 40\%, the polarized intensity is
about 8.8--13.3~mK, which is consistent with the zero-level restored
observations~ (Fig.~\ref{6cmpipa}, right panel).

\subsubsection{Depolarization}

Depolarization can happen in three ways \citep{bur66,tri91,sbs+98},
depth depolarization, beam depolarization and bandwidth
depolarization.  All of them are related with the rotation measure
($RM$) and its variation $\sigma_{\rm RM}$. For depth depolarization,
thermal electrons and relativistic electrons coexist within the same
volume. Radio emission originating from different locations along the
line of sight may have different polarization angles due to Faraday
rotation, and the sum of all these Faraday rotated emission components
along one line of sight will reduce the amount of observed
polarization to some extent. The depolarization can be written as
$DP=|\frac{1-\exp(-S)}{S}|$, where
$S=2\sigma_{RM}^2\lambda^4-2i\lambda^2\mathcal R$. Here $DP$ is
defined as the ratio of the observed polarized intensity to the
intrinsic polarized intensity, and $\mathcal R$ is the RM through the
entire source.

Beam depolarization occurs when the polarization angles vary across
the beam, and therefore the average of the polarized emission within
one beam will result in depolarization. Note that this transverse
variation of $RM$ can happen both in the emission region and in the
foreground medium. For the foreground case, the depolarization can be
written as $DP=\exp(-2\sigma_{RM}^2\lambda^4)$.  Note that the $RM$
varies along the line of sight for depth depolarization
but transversely to the line of sight for the beam depolarization.

The bandwidth depolarization happens when the polarization angles of
the observed emission rotate significantly within the bandwidth. The
depolarization $DP={\rm sinc}(2RM\lambda^2\frac{\Delta\nu}{\nu})$,
where $\Delta\nu$ is the bandwidth of the receiving system.  For our
$\lambda$6~cm observations, although the bandwidth is 600~MHz, a $RM$
of around 3000~rad~m$^{-2}$ is needed for total depolarization, so we
consider bandwidth depolarization as not important for our survey
field.

\subsubsection{Faraday screens}

There are lots of clumps of warm ionized medium in our
Galaxy. These clumps do not emit polarization, but they impose
Faraday effects on the polarization from behind. Low-density ionized gas  
does not show enhanced H$\alpha$ emission, but may locally show a
remarkably strong regular magnetic field, as we see from the Faraday
screen G125.6$-$1.8. The Faraday effects can result in: (1) rotation
of the polarization angle $\psi_{\rm s}=RM_{\rm s}\lambda^2$, where
$RM_{\rm s}$ is the RM of the screen and $\lambda$ is the observation
wavelength; (2) depolarization as it has been described above.  Beside
the diffuse ionized gas, discrete thermal HII regions 
\citep{gdm+01,ulgr03}, possibly HI clouds~\citep{drrf99} and the
surface of molecular clouds \citep{wr04} all can act as a Faraday screens.  In
this paper, the polarized emission originating from behind the Faraday
screen is called "background polarization" and the polarization in front
of the Faraday screen is called "foreground polarization". The
quantities are referred to as ``on'', when the line of sight passes
through the screen and ``off'' otherwise.

The ``on'' components of $U$ and $Q$ data can be represented in 
the following way,
\begin{equation}\label{uqon}
\setlength\arraycolsep{0.2pt}
\left\{
\begin{array}{rcl}
U_{\rm on}&=&PI_{\rm fg}\sin2\psi_{\rm 0,\,fg}+
f PI_{\rm bg}\sin2(\psi_{\rm 0,\,bg}+\psi_{\rm s})\\[2mm]
Q_{\rm on}&=&PI_{\rm fg}\cos2\psi_{\rm 0,\,fg}+
f PI_{\rm bg}\cos2(\psi_{\rm 0,\,bg}+\psi_{\rm s})
\end{array}
\right.
\end{equation}
where ``fg" and ``bg" denote the foreground and background
polarization, $f$ is the depolarization factor at 4.8~GHz, 
$\psi_0$ is the intrinsic polarization angle which does not 
depend on the observation frequency, and $\psi_s$ is the angle 
rotated by the Faraday screen. 

As can be seen from Fig.~\ref{6cmpipa}, the polarization angles tend
to be $0\degr$ in the restored maps, indicating that
$\psi_{0,\,bg}=\psi_{0,\,fg}\approx0\degr$.  For this situation,
$U_{\rm on}$ and $Q_{\rm on}$ can be simplified from Eq.~(\ref{uqon})
as,
\begin{equation}\label{uqon_sim}
\setlength\arraycolsep{0.2pt}
\left\{
\begin{array}{rcl}
U_{\rm on}&=&f PI_{\rm bg}\sin2\psi_{\rm s}\\[2mm]
Q_{\rm on}&=&PI_{\rm fg}+f PI_{\rm bg}\cos2\psi_{\rm s}
\end{array}
\right.
\end{equation}
Then the ``on'' polarized intensity ($PI_{\rm on}$) and polarization angle 
($\psi_{\rm on}$) are calculated as 
$PI_{\rm on}=\sqrt{U^2_{\rm on}+Q^2_{\rm on}}$ and 
$\psi_{\rm on}=\frac{1}{2}{\rm atan}\frac{U_{\rm on}}{Q_{\rm on}}$.
If the line of sight does not pass any Faraday screens, we have the ``off'' 
components as 
$PI_{\rm off}=PI_{\rm fg}+PI_{\rm bg}$ and $\psi_{\rm off}=0\degr$. 
Then the $PI$ ratio ($PI_{\rm on}$/$PI_{\rm off}$) and $PA$ 
difference ($\psi_{\rm on}-\psi_{\rm off}$) is obtained by, 
\begin{equation}\label{pipa_onoff}
\setlength\arraycolsep{0.2pt}
\left\{
\begin{array}{rcl}
\displaystyle{\frac{PI_{\rm on}}{PI_{\rm off}}}&=&
\sqrt{f^2(1-c)^2+c^2+2f c(1-c)\cos2\psi_{\rm s}}\\[2mm]
\psi_{\rm on}-\psi_{\rm off}&=&\displaystyle{\frac{1}{2}{\rm atan}
\frac{f(1-c)\sin2\psi_{\rm s}}{c+f(1-c)\cos2\psi_{\rm s}}}
\end{array}
\right.
\end{equation}
where $c=PI_{\rm fg}/(PI_{\rm fg}+PI_{\rm bg})$ is the fraction of 
foreground polarization.

\begin{figure}
\includegraphics[angle=-90,width=0.46\textwidth]{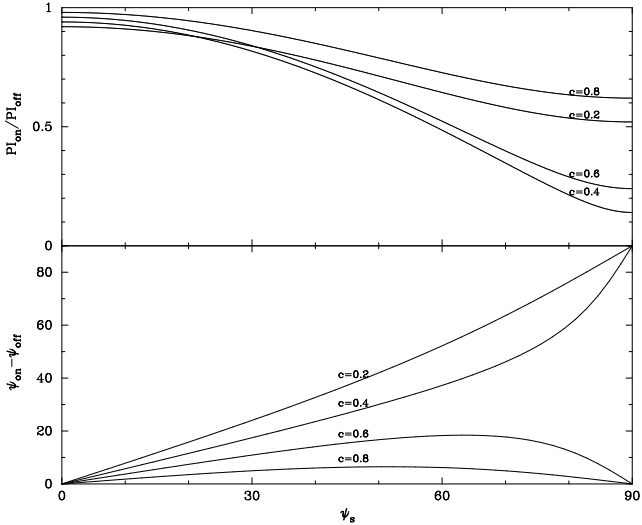}
\caption{The $PI$ ratio ($PI_{\rm on}$/$PI_{\rm off}$) and $PA$ 
difference ($\psi_{\rm on}-\psi_{\rm off}$) varying with the 
angle rotated by the Faraday screen ($\psi_{\rm s}$) are plotted 
for a depolarization of $f=0.9$ and the ratio $c$ between
the foreground and the total PI, $c=0.2,0.4,0.6,0.8$.}
\label{fs_model}
\end{figure}

Based on Eq.~(\ref{pipa_onoff}), the $PI$ ratio and $PA$ difference
are plotted for $f=0.9$ and $c=0.2,0.4,0.6,0.8$ in
Fig.~\ref{fs_model}.  We can see that even with very little
depolarization for the background components, the $PI$ ratio can be
small. We also infer that a small $PA$ difference may be caused by a
large rotation angle of the Faraday screen provided with a large
foreground polarization fraction (larger $c$). All this must be taken
into account in any detailed analysis as discussed below.

 \citet{wr04} have proposed a similar model to fit the relations of
$PI$ with $PA$, the spectral index of polarized intensity with radius
and the $RM$ with radius for Faraday screens identified from the EMLS
survey. In this paper, we use a model to account for the variations of
the $PI$ ratio and the $PA$ difference with radius, which are
independent in respect to the absolute polarization level.

\subsection{Studies of individual objects}
 
\subsubsection{SNRs G126.2$+$1.6 and G127.1$+$0.5}
The SNR G126.2$+$1.6 was discovered by \citet{rks79} based on its
steep spectrum. Later the detection of optical emission lines and the
line ratios confirmed the source as a SNR undoubtedly
\citep{bkg+80,fgk83}.  \citet{frs84} could not rule out a spectral
break at about 3~GHz.  \citet{tl06} suggested a break frequency at
about 1.5~GHz. These uncertainties are created by 
the lack of a precise flux density at $\lambda$6~cm, which was quoted
as an upper limit by \citet{frs84}. As we show below, the spectral
curvature is ruled out using our new $\lambda$6~cm flux density.

SNR G127.1$+$0.5 was suggested to be a SNR by \citet{pau77} and
\citet{cas77}. The central source was originally proposed to be
physically connected with the SNR like the system of SS433 in SNR W50
\citep{cas77,gs82}, but later it was identified as an extragalactic
source by HI absorption observations~ \citep{pgg+82,gg84}. A high
polarization percentage of 25\% at 2695~MHz and 30\% at 4750~MHz were
reported from the Effelsberg data \citep{frs84}, which confirms the
source as a SNR unequivocally. Optical emission from G127.1$+$0.5 was
detected by \citet{xpp+93}.

We obtained the following net flux densities: 2.6$\pm$0.6~Jy for
G126.2$+$1.6 and 6.3$\pm$0.7~Jy for G127.1$+$0.5.  We then revised all
previous measurements by a new extragalactic source correction based
on the NVSS catalog, as listed in Table~\ref{g126g127int}, which
ensures the consistency of the data. We note that one of the central
sources 0125$+$628 in G127.1$+$0.5 is thermal and has a flat spectrum
\citep{jrd89,lt06} with a flux density of 0.368~Jy at 1400~MHz from
the NVSS. For this source, its flux density is taken to be 0.368~Jy at
frequencies larger than 865~MHz and 0.12~Jy at 408~MHz
\citep{jrd89,lt06}.

\begin{table} [!htbp]
\caption{Integrated flux densities of SNRs G126.2$+$1.6 and 
G127.1$+$0.5 after subtraction of compact sources.} 
\label{g126g127int}
\begin{center}
\begin{tabular}{cr@{$\pm$}lcr@{$\pm$}ll}\hline\hline
Frequency &\multicolumn{5}{c}{Flux density (Jy)}&Ref.\\
\cline{2-6}
(MHz) &\multicolumn{2}{c}{G126.2$+$1.6}& &
\multicolumn{2}{c}{G127.1$+$0.5}& \\ \hline
 408 &  9.7 & 3.9 &  & 16.3  & 1.7 &  1,2\\
 408 & 11.5 & 2.5 &  & 16.6  & 2.0 &  3\\
 865 &  5.8 & 1.6 &  & 13.6  & 0.8 &  4\\
1410 &  5.2 & 0.8 &  & 10.2  & 1.2 &  5\\
1420 &  6.7 & 2.1 &  &  9.8  & 0.8 &  1,2\\
1420 &  \multicolumn{2}{c}{} &  &9.7 & 0.8 &  3\\
2695 &  3.9 & 0.4 &  &  7.7  & 0.6 &  5\\
4800 &  2.6 & 0.6 &  &  6.3  & 0.7 &  6\\
4850 &  \multicolumn{2}{c}{} &  & 5.6 & 0.4 &  5\\
\hline
\end{tabular}
\end{center}  
{\footnotesize References: 1 \citet{tl06}; 2 \citet{lt06}; 3 \citet{jrd89}; 
 4 \citet{rzf03}; 5 \citet{frs84}; 6 this paper }
\end{table}

\begin{figure} 
\centering
\includegraphics[angle=-90,width=0.47\textwidth,clip]{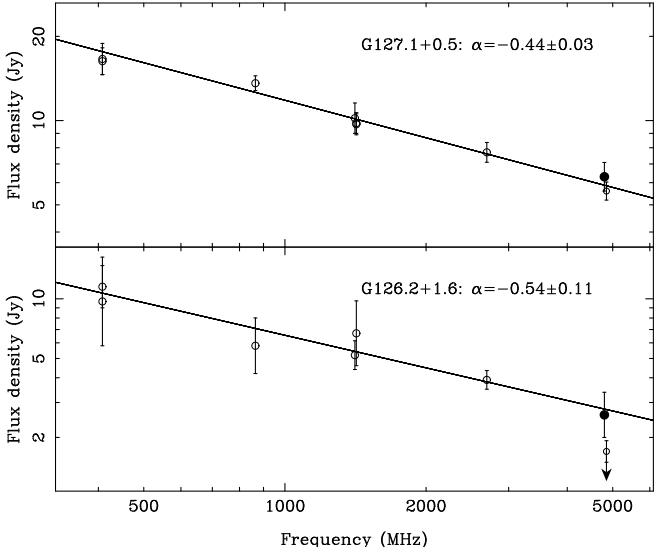}
\caption{The spectra of SNRs G126.2$+$1.6 (lower) and G127.1$+$0.5
(upper). The upper limit of $1.7\pm0.2$~Jy (without source
correction) for G126.2$+$1.6 at 4.85~GHz by \citet{frs84} is also
marked. Our new 4.8~GHz data are indicated by filled circles.}
\label{spec_int}
\end{figure}

\begin{figure}
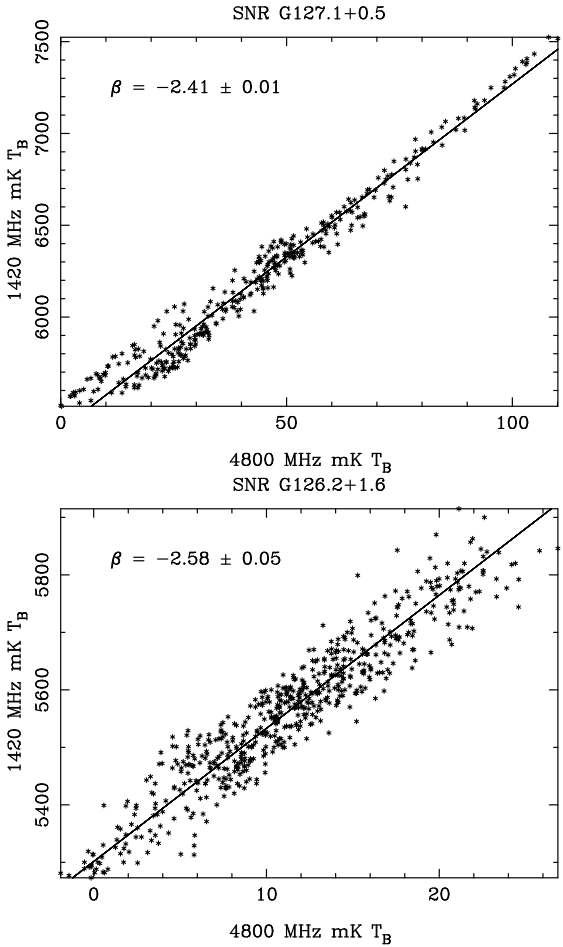
 
\includegraphics[angle=-90,width=0.4\textwidth]{g127.1_4800_1420.ps}
\includegraphics[angle=-90,width=0.4\textwidth]{g126.2_4800_1420.ps}
\caption{TT-plots between the Urumqi 4800~MHz data and the 1420~MHz 
CGPS/Effelsberg data for SNRs G126.2$+$1.6 and G127.1$+$0.5.}\label{tt}
\end{figure}

Using the data listed in Table~\ref{g126g127int}, we fitted the
spectrum for both SNRs as shown in Fig.~\ref{spec_int}.  The linear
fit yields a spectral index of $-$0.54$\pm$0.11 for G126.2$+$1.6 and
$-$0.44$\pm$0.03 for G127.1$+$0.5. It can be clearly seen from
Fig.~\ref{spec_int} (upper panel) that there is no indication of a
spectral break in the frequency range from 408~MHz to 4800~MHz. The
previous suggestion for a spectral curvature \citep{frs84,tl06} is
clearly due to the lack of a precise flux density measurement at
$\lambda$6~cm.

We also checked the TT-plots between the Urumqi 4800~MHz data and the
data at other frequencies. As an example, the TT-plots between our
4800~MHz data and the CGPS/Effelsberg 1420~MHz data for both SNRs are
shown in Fig.~\ref{tt}.  As can be seen from Table~\ref{tt_par}, for
SNR G127.1$+$0.5 the weighted average $\alpha$ is $-$0.41$\pm$0.02 and
$-$0.52$\pm$0.03 for SNR G126.2$+$1.6, both agree well with the
spectral indices derived from the integrated flux densities.

Fig.~\ref{6cmpipa} shows polarization towards both SNRs in the
original and the zero-level restored map.  SNR G127.1$+$0.5 shows strong
polarization, much larger than the Galactic contribution, and the
restoration process introduces just slight changes.  Polarization
B-vectors follow the shell, conforming to the model by
\citet{laa62}. The orientation of the symmetric axis of magnetic
fields is parallel to the local large-scale magnetic field, which
supports the idea proposed of a barrel type SNRs by \citet{fr90}.
This also shows the importance of the magnetic field in shaping the
evolution of SNRs.

For SNR G126.2$+$1.6 the polarization intensity is quite weak towards
the eastern shell compared to its surroundings,
and the orientation of the B-vectors is nearly perpendicular to the
shell. Since SNR G126.2$+$1.6 is fully evolved \citep{tl06}, this
magnetic field configuration is considered as rather atypical
\citep{fr04}. This must be ascribed to the influence of the foreground
or background polarization. Due to the observed $U_{\rm obs}\approx0$
and $U_{\rm bg}=U_{\rm fg}\approx0$, together with
Eq.~(\ref{uqon_sim}) we can obtain $\psi_s=0\degr$.  Thus the observed
$Q_{\rm obs}=Q_{\rm bg}+Q_{\rm fg}+Q_{\rm SNR}$.  In the maps of
original $U$ and $Q$~(Fig.~\ref{6cmuq}), we see that $U_{\rm
SNR}\approx0$ and $Q_{\rm SNR}<0$. However for the background or
foreground polarization, we have $U_{\rm fg/bg}\approx0$ and $Q_{\rm
fg/bg}>0$. So the addition of all the polarization will partly cancel
the $Q_{\rm obs}$. Therefore the polarized intensity might get
reduced, while the polarization angle is still near $0\degr$.

\subsubsection{G124.9$+$0.1: a newly identified HII region}

\begin{figure}
\includegraphics[bb=63 37 635 774,clip,angle=-90,width=0.47\textwidth]{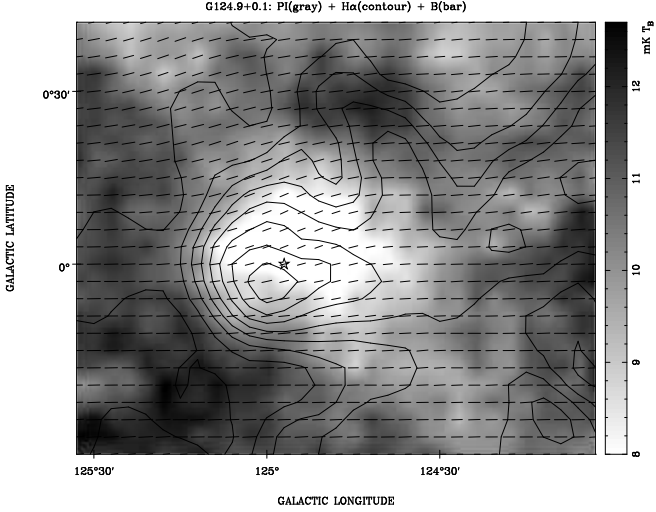}
\caption{HII region G124.9$+$0.1: polarized intensity is 
gray-scale coded, H$\alpha$ intensity is shown by contours, and
the orientation of the magnetic field in given by the direction of
bars. The H$\alpha$ contours start at 8~Rayleigh and run in steps of 2~Rayleigh. The
position of the B0~III star Hilt 102 is indicated.}
\label{pi_g124.9}
\end{figure}

G124.9$+$0.1 is an extended source with a radius of about $27\arcmin$
and a flux density of 1.4$\pm$0.2~Jy (Fig.~\ref{pi_g124.9}).  In the
original polarization map (Fig.~\ref{6cmpipa}), weak but distinct
polarization across this extended source gives a hint for a possible
SNR. However, after the restoration, a hole in polarized intensity
becomes obvious, which is caused by a Faraday screen. Polarization
angles inside the source deviate from those of its surroundings. The
known HII~region Sh~186 is located near its northern edge.

The average spectral index of G124.9$+$0.1 obtained by a TT-plot is
$\alpha=0.004$, confirming its thermal nature. The ratio of 60~$\mu$m
flux density to 6~cm flux density (Fig.~\ref{060_6cm} and
Table~\ref{tt_par}) is 1100 and consistent with HII~region properties.
We find a B0~III star Hilt 102 at $(l, b)=(124\fdg95, -0\fdg01)$,
which could be the exciting star for G124.9$+$0.1 (see
Fig.~\ref{pi_g124.9}).  The distance modulus of the star is 12.2~mag,
what corresponds to a distance of about 2.8~kpc. We conclude that
G124.9$+$0.1 is a newly identified HII region.

Thermal emission from HII~regions is generated by free-free emission
of electrons. The observed radio continuum emission at 4.8~GHz
requires a number of ionizing photons $N_{LC}$ \citep{rub68,bmd+97} as
given by:
\begin{equation}\label{nlc}
N_{LC}=8.825\times10^{46}T_4^{-0.45}SD^2
\end{equation}
where $N_{LC}$ is in units of s$^{-1}$, $T_4$ is the electron
temperature in 10$^4$~K, $S$ is the flux density in Jy and $D$ is the
distance in kpc. Based on the radio continuum measurements at 4.8~GHz,
the average electron density can be obtained by assuming a spherical
model \citep{mh67},
\begin{equation}\label{ne}
n_e=949T_4^{0.175}S^{0.5}D^{-0.5}\theta^{-1.5}
\end{equation}
where $n_e$ is in units of cm$^{-3}$, $\theta$ is the apparent
diameter of the source in arcmin and the other parameters are as in
Eq.~(\ref{nlc}). The electron density could also be derived from the
emission measure $EM$ defined as the integral of the square of
electron density through the source along the line of sight.
The $EM$ is related to the H$\alpha$ intensity
as~\citep{hrt98},
\begin{equation}\label{em}
EM=2.75T_4^{0.9}I_{\rm H\alpha}\exp\left[2.44E(B-V)\right]
\end{equation}   
where $EM$ is in units of pc~cm$^{-6}$, $I_{\rm H\alpha}$ is the
H$\alpha$ intensity in Rayleigh, and $E(B-V)$ is the reddening in
magnitudes.

Assuming an electron temperature of 8000~K for G124.9$+$0.1 we obtain
the required $N_{LC}$ of 1.07$\times10^{48}$~s$^{-1}$, which conforms
to the value given in \citet{pan73} for a B0~III star. According to
Eq.~(\ref{ne}) we derive an electron density of 1.6~cm$^{-3}$.

\begin{figure}
\includegraphics[angle=-90,width=0.48\textwidth]{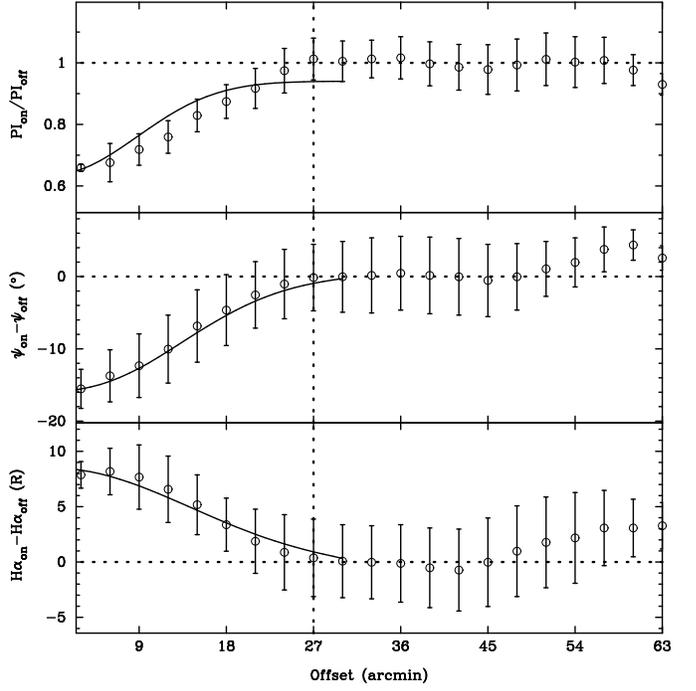}
\caption{Variation of the difference of the H$\alpha$ intensity
between ``on'' and ``off'' positions, $PA$ difference and $PI$ ratio
from the center of the HII~region at $(l,b)=(124\fdg90, 0\fdg05)$ to
the outskirts.  The vertical dotted line marks the boundary for the
``on'' and ``off'' regions and also indicates the size of the
HII~region.  The solid lines give a fit according to the Faraday
screen model discussed in text.}
\label{dp_g124.9}
\end{figure}

We overlaid the H$\alpha$ data on the restored 4.8~GHz polarization data 
in Fig.~\ref{pi_g124.9}. The H$\alpha$ data
are taken from the all-sky H$\alpha$ template by \citet{fin03}.  As
can be seen from Fig.~\ref{pi_g124.9}, there is a clear
anti-correlation between the polarization intensity and the H$\alpha$
intensity. The polarization angles rotate where the H$\alpha$ emission
is strong. To show their correlations, we plotted the radial
distribution of these quantities. The center is selected to be located
at $(l, b)=(124\fdg90, 0\fdg05)$, the quantities are averaged within
rings of $3\arcmin$-width starting from the center. According to
Fig.~\ref{dp_g124.9}, we take the radius of $27\arcmin$ as the size of
the HII~region. Within the source the $PA$ difference and $PI$ ratio
increase gradually, while the H$\alpha$ intensity difference decreases
towards larger radii. The $PI$ ratio varies from about 0.66 to nearly
1, and the $PA$ difference varies from about $-16\degr$ to about
$0\degr$.

We first investigate the difference of the H$\alpha$ intensity, which
gives us hints for the variation of the electron density and the path
length, because $I_{\rm H\alpha}\propto EM= n_e^2 l$. Here $l$ is the
length of the line of sight within the source. A Gaussian electron
density distribution \citep[e.g.][]{mh67} and a pathlength within a
spheroid could account for the observation in
Fig.~\ref{dp_g124.9}. The pathlength calculates as
$l(\phi)=l_0\sqrt{1-\frac{D^2}{R^2}\phi^2}$, where $l_0=2R$ is the
maximal pathlength passing the center, $\phi$ the offset from the
center, $R$ the radius and $D$ the distance. Then the H$\alpha$
intensity can be written as $I_{\rm
H\alpha}(\phi)=I_0e^{-\frac{\phi^2}{\sigma^2_{\rm H\alpha}}}
\sqrt{1-\frac{D^2}{R^2}\phi^2}$ and the best fitting parameters are
$I_0=8.5$~Rayleigh and $\sigma_{\rm H\alpha}=20\arcmin$. According to
Eq.~(\ref{em}), the maximal H$\alpha$ intensity of 8.5~Rayleigh
corresponds to an $EM$ of 260~pc~cm$^{-6}$ with a reddening $E(B-V)$
of 1.07 \citep{hil56}. Since the $EM$ can be estimated as $2n_e^2R$,
$n_e$ is derived as 2.3~cm$^{-3}$, which is fairly consistent with a
density of 1.6~cm$^{-3}$ as estimated before.

We interpret the observed $PA$ difference and $PI$ ratio using the
Faraday screen model in Eq.~(\ref{pipa_onoff}). Based on the H$\alpha$
intensity fit, we model the $PA$ rotation imposed by the Faraday
screen as $\psi_{\rm s}(\phi)=\psi_0e^{-\frac{\phi^2}{\sigma^2_{\rm
s}}} \sqrt{1-\frac{D^2}{R^2}\phi^2}$. Here $|\psi_0|$ is the maximal
rotation angle. The best parameters for a reasonable fit for both the
$PI$ ratio and $PA$ difference are $\psi_0=-50\degr$, $f=0.85$,
$c=0.6$ and $\sigma_{\rm s}=18\arcmin$. The maximal rotation angle of
$50\degr$ at 4.8~GHz corresponds to a maximal $RM$ of
223~rad~m$^{-2}$. The maximal RM can be estimated as
$RM=2Kn_eB_{||}R$, where $K = 0.81$ is a constant.  With a distance of
about 2.8~kpc and a radius of 27\arcmin, we obtain about 22~pc for
$R$.  $n_e$ was taken to be 1.6~cm$^{-3}$. We obtain a magnetic field
component along the line of sight of $B_{||}\approx 3.9~\mu$G, which
is consistent with the magnetic field strength derived for other
HII~regions based on excessive RMs of extragalactic sources observed
in their direction \citep{hct81}.

As can be seen from Fig.~\ref{dp_g124.9}, the fit for the $PI$ ratio
is roughly fine but not exact. A depolarization factor $f\approx1$ is
needed to fit solely the $PI$ ratio, which means that the Faraday
screen does not depolarize the background polarization.  It could be
that there exist an extended envelope around the HII region, which
causes an underestimation of the rotation angle ($\psi_{\rm s}$) of
the screen by several degrees. But this does not affect the $PI$
ratio, which is nearly constant for small $\psi_{\rm s}$ as can be
seen from Fig.~\ref{fs_model}.

The fitting above also gives a foreground polarization of about 7~mK,
which allows us estimate the emissivity of the synchrotron emission in
this direction.  For a distance of 2.8~kpc we calculate an emissivity
for the polarized intensity of 2.5~mK~kpc$^{-1}$. Assuming a
polarization percentage of about 40\%, we got the emissivity for total
intensity of 6.3~mK~kpc$^{-1}$ at 4.8~GHz, corresponding to an
emissivity at 22~MHz of about 22~K~pc$^{-1}$ for a spectral index of
$-$2.8 and about 7.6~K~pc$^{-1}$ for a spectral index of $-$2.6.
\citet{rcls99} reported a synchrotron emissivity of 20.9~K~pc$^{-1}$
obtained using the absorption towards the nearby HII region IC~1805
with a distance of 2.2~kpc located at $(l,b)=(134\fdg8, 0\fdg9)$ at
22~MHz. All results are consistent with a spectral index of $-$2.8.

\subsubsection{An extended shell emerging from Sh~185}

The nebula Sh 185 contains two reflection nebulae IC~63 and IC~59
\citep{bmd+97}, which could be marginally resolved by our
observation. The total flux density of Sh~185 at 6~cm is
1.4$\pm$0.5~Jy. The complex is illuminated by the B0~IV star
$\gamma$~Cas at a distance of 190~pc \citep{plk+97}.

\begin{figure}[htbp]
\includegraphics[width=0.5\textwidth]{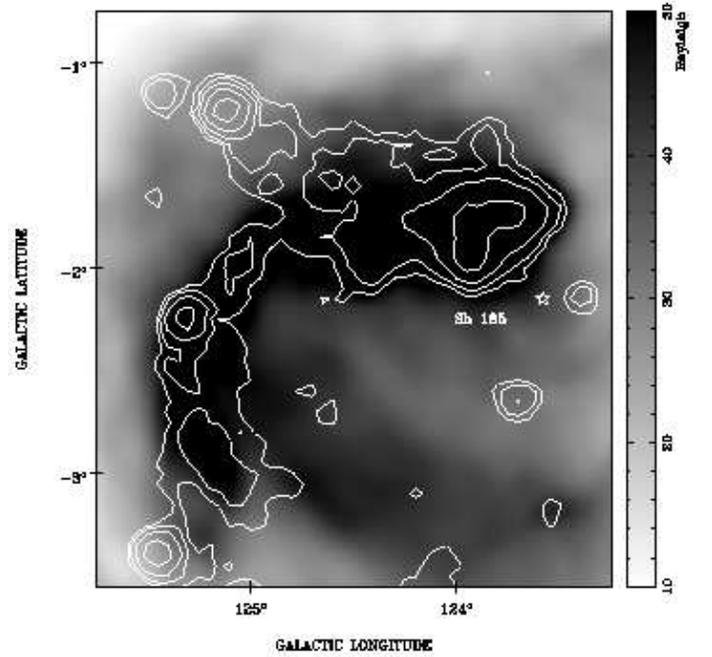}
\caption{The total intensity at 6~cm of Sh~185 is shown in contours
with the same (positive) levels as in Fig.~\ref{6cmi}. The three point
sources to the left are probably background extragalactic radio
sources.  The gray-scale image shows the H$\alpha$ intensity. The
marked star indicates the position of $\gamma$~Cas.}
\label{sh185_shell}
\end{figure}

The most spectacular new feature we can discern from our total
intensity map at 6~cm is a shell towards south-east of
Sh~185, which exactly follows the H$\alpha$ shell
(Fig.~\ref{sh185_shell}). The radius of the Str\"omgren sphere of
$\gamma$~Cas is 6.5~pc \citep{knm05} and therefore the shell is
probably illuminated by $\gamma$~Cas.  Since the shell is very weak,
we could only get the upper limit of about 0.5~Jy for the integrated
flux density, which requires ionizing photons of
$1.76\times10^{45}$~s$^{-1}$ for an assumed temperature of the shell
of 8000~K.  The total ionizing photons from $\gamma$~Cas is
$(5.5\pm1.5)$$\times10^{46}$~s$^{-1}$ \citep{bmd+97}.  From the star,
the shell subtends an angle of about
$25\degr$~(Fig.~\ref{sh185_shell}) and therefore receives the photons
of about $3.85\times10^{45}$~s$^{-1}$ from the star. This is
sufficient to maintain the continuum radiation from the shell.

Sh~185 as well as the shell does not 
have any measureable effect at $\lambda$6~cm on the polarization observations. 

\begin{figure}[hbt]
\includegraphics[angle=-90,width=0.45\textwidth]{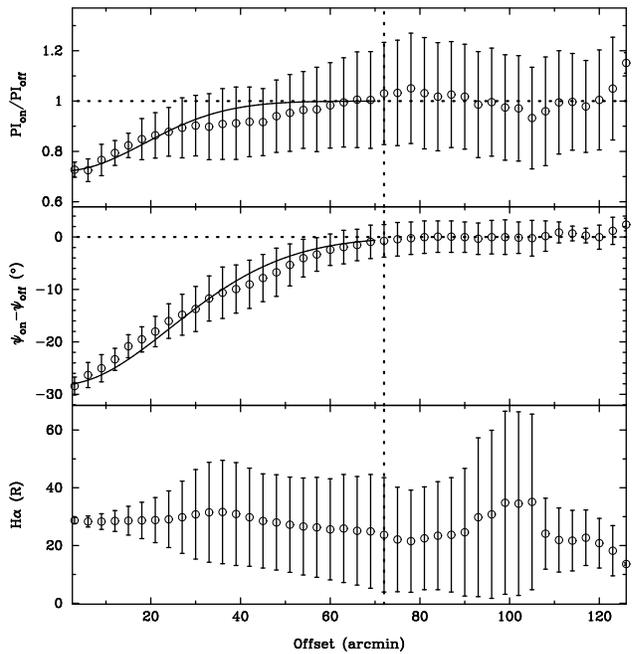}
\caption{The same ring integrated profiles as shown in Fig.~\ref{dp_g124.9}, 
but for G125.6$-$1.8. }
\label{dp_fs}
\end{figure}

\subsubsection{G125.6$-$1.8: a large Faraday screen}

To the east of Sh~185 and its eastern shell, there is an outstanding
region with excessive polarization centered at $(l,b)=(125\fdg6,-1\fdg8)$ 
on the original Urumqi polarization map (Fig.~\ref{6cmpipa}) and slightly 
reduced polarization after the restoration. The
polarization angles are significantly modulated when compared to the
surrounding angles and there is no correspondence in the total
intensity map. This region has an apparent size of about 2\fdg4 when
fitted with a spheroid. The largest $PA$ difference at the center is
about $-30\degr$, but steadily decreases towards the edge.  The $PI$
ratio is nearly constant or about 90\% as seen in Fig.~\ref{dp_fs},
which means neither depolarization nor foreground components, i.e.
$f\approx1$. Similar to G124.9$+$0.1, we also fit the observed $PI$
ratio and $PA$ difference with the Faraday screen model in
Eq.~(\ref{pipa_onoff}), and got following parameters:
$\psi_0=-45\degr$, $\sigma_{\rm s}=36\arcmin$, $f=1$ and $c=0.4$.  The
maximal rotation from the Faraday screen is about $45\degr$
corresponding to a RM of about 200~rad~m$^{-2}$.

The fraction of foreground polarization of 40\% corresponds to a
brightness temperature of about 3.6~mK. If we take the polarized 
emissivity of 2.5~mK~kpc$^{-1}$ determined in the direction of G124.9$+$0.1, 
we obtain a screen distance of about 1.4~kpc. In case we use the 
synchrotron emissivity of 11~mK~kpc$^{-1}$ extrapolated from that quoted
be \citet{bkb85} at 408~MHz for a spectral index of $-2.8$ and a 
polarization percentage of 
40\%, we obtain a distance of about 0.8~kpc. For further calculations we 
take the average of both distance estimates, 
about 1.1~kpc, and get about 46~pc for the screen size. The large 
distance also excludes the possibility that the screen is related to the 
local complex Sh~185, although the Faraday screen partly overlaps with 
the northeastern part of the Sh~185 complex. 

Towards the screen, the H$\alpha$ intensity is about 30~Rayleigh 
(Fig.~\ref{dp_fs}), which is predominantly attributed to the Sh~185 nebula. 
There is no indication of excessive H$\alpha$ associated to the screen at 
all. However, the H$\alpha$ 
emission from the screen might be totally masked by the nebula Sh~185. 
The H$\alpha$ absorption is very uncertain in the Galactic plane, therefore 
we cannot infer an electron density from the H$\alpha$ results.  

We see no signature from this screen in the total intensity map. This means the 
total intensity of this object must be less than about five times the noise 
or 4.25~mK. The brightness temperature contributed by 
the screen can be represented as $T_s=\tau T_e=6n_e^2$~mK, where
the opacity $\tau$ is calculated as in \citet{rw00} and the electron 
temperature $T_e$ is taken as 8000~K. Since $T_s<4.25$~mK, we obtain the 
upper limit of the electron density as 0.84~cm$^{-3}$, assuming a spherical
shape for the Faraday screen. This electron density corresponds to an
upper limit for the 
H$\alpha$ intensity of about 1.3~Rayleigh for an $E(B-V)$ lower 
limit of 1~mag near the Galactic plane. This is beyond the detection limit 
in the all-sky H$\alpha$ map by \citet{fin03}. Together with the $RM$ 
derived, we obtain a lower limit for the regular magnetic field along the 
line of sight of 6.4~$\mu$G. This is smaller than the magnetic field 
strengths derived by \citet{wr04} for a number of small Faraday screens 
having $\sim$2~pc in size, which are located at the edge of local molecular 
clouds in Taurus. However, the line of sight magnetic field component we 
find for G125.6$-$1.8 clearly excceds the average {\it total} field strength 
in the interstellar medium at this Galactocentric distance \citep{hml+06}. 
Faraday screens due to a thermal 
electron density excess and avoiding an enhanced regular magnetic field 
component have been discussed by \citet{gldt98} and \citet{ul02}. In these 
two cases the distances for the objects could be not be well constrained 
and make such an interpretation possible. In the case of G125.6$-$1.8 an 
enhanced magnetic field strength seems unavoidable to be compatible with 
the available data. 

\subsubsection{Sh~183}
\citet{larv92} studied the radio morphology of the HII~region Sh~183
using the DRAO array at 408~MHz and 1420~MHz.  They proposed an so far
unknown O~5.5 star for its excitation. Unfortunately we can not find
any O- or B-type star towards Sh~183. 
The recombination and HI velocity of Sh~183 is about $-$63~km~s$^{-1}$
 \citep{larv92}.
This means a dynamical distance of about 6~kpc with the solar 
parameters $R_0=8.5$~kpc and $\Theta_0=220$~km~s$^{-1}$. However 
due to the spiral shock at the leading edge of the Perseus arm \citep{rob72},
the dynamical distance has a large error. Measurements of Cas~OB7 
show that the radial velocity of the gas can be displaced by about 
$-$20~km~s$^{-1}$ due to that shock \citep{cp03}. Since Sh~183 is located at 
the boundary 
of Cas OB7, we make a similar correction and obtain a distance of about 
3.6~kpc, which places Sh~183 on the far side or just behind the Perseus arm. 

We determined a flux density for
this HII~region at 4.8~GHz of 5.3$\pm$0.5~Jy, which yields a spectral
index of $-$0.05$\pm$0.05 when combined with flux densities at lower
frequencies \citep{larv92}. This spectral index is confirmed using the
TT-plot method (Table~\ref{tt_par}).

\begin{figure}
\includegraphics[width=0.5\textwidth]{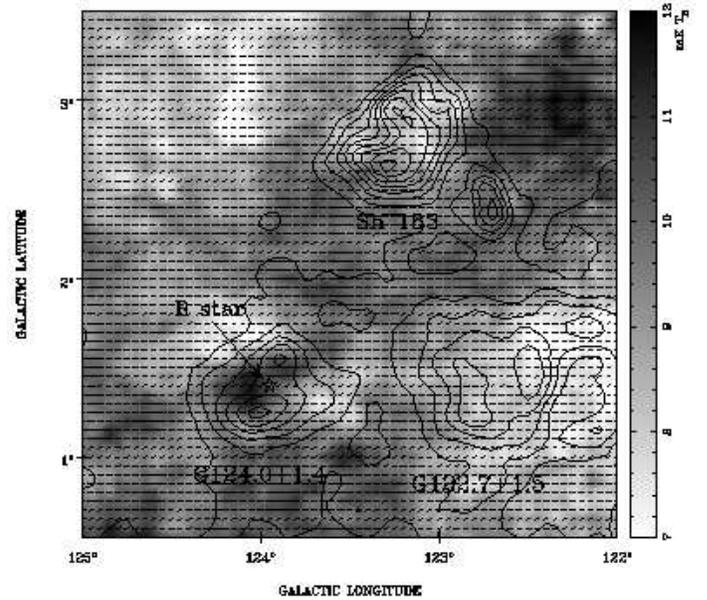}
\caption{The same as Fig.~\ref{pi_g124.9} but for the region
encompassing the extended ojbects Sh~183, G124.0$+$1.4 and G122.7+1.2.
The H$\alpha$ intensity contours start from 10~Rayleigh and run in
steps of 3~Rayleigh. A B-type star located in the center of
G124.0$+$1.4 is marked.  }
\label{dp_s183}
\end{figure}


Towards the direction of the HII region Sh~183, some weak polarization
minima are visible (Fig.~\ref{dp_s183}). However, the polarization
angles do not differ from its surroundings, which means that the
polarized emission predominately originates in the foreground in respect to 
Sh~183. Considering together the fact that towards the Faraday screen 
G124.9$+$0.1 about 80\% of the polarized emission is originated within the 
distance of 3.6~kpc and other screen, G125.6$-$1.8, all polarized emission 
is originated within a distance of about 3~kpc, we conclude that the 
polarization emission originated behind the Perseus arm is very weak. 

\subsubsection{Possible new faint HII~regions}

The spectral index of the extended source G124.0$+$1.4 is about
$-$0.22 from the TT-plots. We found a B-type star near the center of
G124.0$+$1.4 (Fig.~\ref{dp_s183}), which means that the source is
likely a HII~region. Unfortunately no further information is available
for this star. The ratio of the IRAS 60~$\mu$m flux density to 4.8~GHz
density (Fig.~\ref{060_6cm} and Table~\ref{tt_par}) is 660, which also
suggests that this source is more likely a HII~region rather than a
SNR.

Other optically identified HII~regions are either too weak at 6~cm
(such as Sh~181 and Sh~180) or confused with complex emission regions
(such as DU~65 and Sh~186), which prevents us from a more detailed
study.

\section{Conclusions}
In this paper we report on the observation strategy and data
processing procedures of the Sino-German $\lambda$6~cm continuum and
polarization survey of the Galactic plane carried out with the Urumqi 25~m 
telescope. Preliminary results for the first survey region
centered at $(l,b)=(125\fdg5, 0\degr)$ are presented.

The maps show many features in both total intensity and polarized
intensity.  From the total intensity maps: (1) The $\lambda$6~cm flux
density of 2.6$\pm$0.6~Jy for SNR G126.2$+$1.6 was measured, which,
together with previous data, rules out a spectral curvature. The
integrated flux densities for other extended sources were also
obtained.  (2) We identified the HII~region G124.9$+$0.1, being
studied in some detail, and G124.0$+$1.4 being most likely also a
HII-region.  (3) A large thermal shell is recognized to be probably
physically connected with the reflection nebula Sh~185 and being
illuminated by the star $\gamma$~Cas.

We tried to compensate the so far missing large-scale structures in
our $\lambda$6~cm $U$ and $Q$ maps by extrapolating WMAP K-band
polarization data from 22.8~GHz to 4.8~GHz with an assumed spectral
index of $-$2.8. This is to be regarded as a trial for 
the zero-level restoration.
The polarized structures change dramatically for some
regions after {zero-level restoration}.  The polarization angles are very
regular in general and trace a uniform large-scale magnetic field
parallel to the Galactic plane. Some Faraday screens which show
enhanced polarization in the original polarized maps are recognized
after zero-level restoration. Based on model fitting for the Faraday 
screens,
the foreground and background polarization could be separated, which
allows us to assess the emissivity of the synchrotron emission.  The
prominent features in the polarization maps are summarized as: (1) The
HII region G124.9$+$0.1 was identified as a Faraday screen in the
zero-level restored maps. Its thermal electron density is estimated to be
about 1.6~cm$^{-3}$ and its magnetic field component along the line of
sight is about 3.9~$\mu$G. (2) Polarization is detected from SNRs
G127.1$+$0.5 and G126.2$+$1.6. For the barrel type SNR G127.1$+$0.5,
the magnetic field fits the direction of the Galactic large-scale
magnetic field. (3) A large diameter Faraday screen G125.6$-$1.8 was
identified, whose electron thermal density is estimated to be up to
0.84~cm$^{-3}$ and its magnetic field of 6.4~$\mu$G or larger.

In summary, the observations of the first region of the $\lambda$6\ cm
survey of the Galactic plane demonstrate their potential to detect
numerous Galactic structures and to reveal new Faraday screens up to
large distances, which are remarkable ISM features with strong and
regular magnetic field.
  
\begin{acknowledgements}
The $\lambda$6\ cm data were obtained with the receiver system from
the MPIfR mounted at the Nanshan 25~m telescope at the Urumqi
Observatory of NAOC. We thank the staff of the Urumqi Observatory of
NAOC for the great assistance during the installation of the receiver
and the observations. In particular we like to thank Otmar Lochner for
the construction of the $\lambda$6\ cm system and its installation and
Maozheng Chen and Jun Ma for their help during the installation of the
receiver and its maintenance. We are very grateful to Dr. Peter
M\"uller for the software needed to make mapping observations and data
reduction. The MPG and the NAOC supported the construction of the
Urumqi $\lambda$6\ cm receiving system by special funds.  We thank
Dr. Roland Kothes for his help on handling CGPS data. The survey team
is supported by the National Natural Science foundation of China
(10473015, 10521001), and the Partner group of the MPIfR at NAOC in
the frame of the exchange program between MPG and CAS for many
bilateral visits. We like to thank the referee Dr. Tom Landecker for 
his helpful comments which significantly improve the paper.
\end{acknowledgements}

\bibliographystyle{aa}
\end{document}